\documentclass{aa}

\usepackage{graphicx}
\usepackage{txfonts}
\usepackage{amsmath}	
\usepackage{upgreek}
\usepackage{xspace}
\usepackage{xcolor}
\usepackage{enumitem}
\usepackage{threeparttable}

\usepackage[colorlinks=true,citecolor=blue, urlcolor=blue, linkcolor=blue]{hyperref}

\makeatletter
\renewcommand*\aa@pageof{, page \thepage{} of \pageref*{LastPage}}
\makeatother

\renewcommand{\arraystretch}{1.4}

\newcommand{\Mp}{\ensuremath{M_{\mathrm{p}}}\xspace}
\newcommand{\Sunits}{\ensuremath{k_{\mathrm{B}}/\mathrm{baryon}}\xspace}
\newcommand{\mum}{\ensuremath{\upmu}m\xspace}
\newcommand{\mas}{\ensuremath{\upmu}as\xspace}

\begin{document}

   \title{Differentiating Formation Models with New Dynamical Masses for the PDS 70 Protoplanets}

   \author{David~Trevascus\inst{\ref{mpia},\ref{Heidelberg}},
        Sarah~Blunt\inst{\ref{CIERA}},
        Valentin~Christiaens\inst{\ref{KUL},\ref{STAR}},
        Elisabeth~Matthews\inst{\ref{mpia}},
        Iain~Hammond\inst{\ref{Monash}},
        Wolfgang~Brandner\inst{\ref{mpia}},
        Jason~Wang\inst{\ref{CIERA}},
        Sylvestre~Lacour\inst{\ref{LESIA},\ref{ESO}},
        Arthur~Vigan\inst{\ref{LAM}},
        William~O.~Balmer\inst{\ref{JHU},\ref{STScI}},
        Mickael~Bonnefoy\inst{\ref{IPAG}},
        Remo~Burn\inst{\ref{mpia}},
        Ga\"el~Chauvin\inst{\ref{mpia},\ref{OCA}},
        Raffaele~Gratton\inst{\ref{INAF}},
        Mathis~Houllé\inst{\ref{IPAG},\ref{OCA}},
        Sasha~Hinkley\inst{\ref{Exeter}},
        Jens~Kammerer\inst{\ref{ESO}},
        Laura~Kreidberg\inst{\ref{mpia}},
        Gabriel-Dominique~Marleau\inst{\ref{mpia},\ref{UDE},\ref{Bern}},
        Dino~Mesa\inst{\ref{INAF}},
        Gilles Otten\inst{\ref{ASIAA}},
        Mathias~Nowak\inst{\ref{LESIA}},
        Emily~Rickman\inst{\ref{ESA-STScI}},
        Joel~Sanchez-Bermudez\inst{\ref{ASU}},
        Jonas~Sauter\inst{\ref{mpia},\ref{Heidelberg}}
        }
    
    \institute{Max-Planck-Institut f\"ur Astronomie, K\"onigstuhl 17, 69117 Heidelberg, Germany
        \label{mpia}
        \and
        Universität Heidelberg, Grabengasse 1, 69117 Heidelberg, Germany
        \label{Heidelberg}
        \and
        Center for Interdisciplinary Exploration and Research in Astrophysics (CIERA) and Department of Physics and Astronomy, Northwestern University, Evanston, IL 60208, USA
        \label{CIERA}
        \and
        Institute of Astronomy, KU Leuven, Celestijnenlaan 200D, Leuven, Belgium
        \label{KUL}
        \and
        Space sciences, Technologies \& Astrophysics Research (STAR) Institute, Universit\'e de Li\`ege, All\'ee du Six Ao\^ut 19c, B-4000 Sart Tilman, Belgium
        \label{STAR}
        \and
        School of Physics and Astronomy, Monash University, Clayton, VIC 3800, Australia
        \label{Monash}
        \and
        LESIA, Observatoire de Paris, Universit\'e PSL, CNRS, Sorbonne Universit\'e, Universit\'e Paris Cit\'e, 5 place Jules Janssen, 92195 Meudon, France 
        \label{LESIA}
        \and
        European Southern Observatory (ESO), Karl-Schwarzschild-Stra\ss{}e~2, 85748 Garching, Germany 
        \label{ESO}
        \and
        Universit\'e Aix-Marseille, CNRS, CNES, LAM, Marseille, France 
        \label{LAM}
        \and
        Department of Physics \& Astronomy, Johns Hopkins University, 3400 N. Charles Street, Baltimore, MD 21218, USA 
        \label{JHU}
        \and Space Telescope Science Institute (STScI), 3700 San Martin Drive, Baltimore, MD 21218, USA 
        \label{STScI}
        \and
        Universit\'e Grenoble Alpes, CNRS, IPAG, 38000 Grenoble, France 
        \label{IPAG}
        \and
        Laboratoire Lagrange, Universit\'e C\^ote d'Azur, CNRS, Observatoire de la C\^ote d'Azur, 06304 Nice, France 
        \label{OCA}
        \and
        INAF -- Osservatorio Astronomico di Padova, Vicolo dell'Osservatorio 5, 35122 Padova, Italy 
        \label{INAF}
        \and
        University of Exeter, Physics Building, Stocker Road, Exeter EX4 4QL, UK
        \label{Exeter}
        \and
        Fakult\"at für Physik, Universit\"at Duisburg--Essen, Lotharstra\ss{}e 1, 47057 Duisburg, Germany
        \label{UDE}
        \and
        Physikalisches Institut, Universit\"at Bern, Gesellschaftsstr.~6, 3012 Bern, Switzerland
        \label{Bern}
        \and
        Academia Sinica, Institute of Astronomy and Astrophysics, 11F Astronomy-Mathematics Building, NTU/AS campus, No. 1, Section 4, Roosevelt Rd., Taipei 10617, Taiwan
        \label{ASIAA}
        \and
        European Space Agency (ESA), ESA Office, STScI, 3700 San Martin Drive, Baltimore, MD 21218, USA
        \label{ESA-STScI}
        \and
        Instituto de Astronom\'ia, Universidad Nacional Aut\'onoma de M\'exico, Apdo. Postal 70264, Ciudad de M\'exico 04510, Mexico
        \label{ASU}
        }

   \date{Received January 28, 2025; accepted ??}

 
  \abstract
  {
  Hot- and cold-start planet formation models predict differing luminosities for the young, bright planets that direct imaging surveys are most sensitive to. However, precise mass estimates are required to distinguish between these models observationally. The presence of two directly imaged planets, PDS 70 \textit{b} and \textit{c}, in the PDS 70 protoplanetary disk provides us a unique opportunity for dynamical mass measurements, since the masses for these planets are currently poorly constrained. Fitting orbital parameters to new astrometry of these planets, taken with VLTI/GRAVITY in the $K$~band, we find $2\sigma$ dynamical upper mass limits of 4.9 $M_{\rm Jup}$ for \textit{b} and 13.6 $M_{\rm Jup}$ for \textit{c}. Adding astrometry from the newly proposed planet candidate PDS 70 \textit{d} into our model, we determine $2\sigma$ dynamical upper mass limits of 5.3 $M_{\rm Jup}$, 7.5 $M_{\rm Jup}$ and 2.2 $M_{\rm Jup}$ for \textit{b}, \textit{c}, and the candidate \textit{d} respectively. However, $N$-body analysis of the orbits fit in this case suggest that the inclusion of $d$ makes the system unstable. Using the upper mass limits for \textit{b} and \textit{c} we rule out the coldest-start formation models for both planets, calculating minimum post-formation entropies ($S_i$) of 9.5 \Sunits and 8.4 \Sunits respectively. This places PDS 70 \textit{b} and \textit{c} on the growing list of directly-imaged planets inconsistent with cold-start formation.}


    \authorrunning{Trevascus et al.}
    \maketitle
%

\section{Introduction}

The sensitivity of direct-imaging surveys is highly dependent on the assumed model of planet formation. Hot-start \citep[e.g.,][]{Baraffe_2003, Phillips_2020}, warm-start \citep[e.g.,][]{Spiegel_2012, Linder_2019}, and cold-start \citep[e.g.,][]{Marley_2007, Fortney_2008} planet evolution models all assume different post-formation entropies ($S_i$), leading to differing luminosity predictions for young planets $\lessapprox 100$ Myr old, which are the typical targets of these surveys. 

Planet masses are traditionally inferred from evolutionary models \citep[e.g.,][]{Bowler_2016}, but these mass estimates are often poorly constrained and dependent upon the choice of model. However, dynamical mass estimation can provide model-independent planet masses with much stronger constraints, such as for $\beta$ Pic b and c \citep{Brandt_2021a}, HR 8799 e \citep{Brandt_2021b, Zurlo_2022} and AF Lep~b \citep{Balmer_2024}. Moreover, precise dynamical mass estimates, in combination with planet ages and luminosities, can be used to constrain $S_i$ for these planets and therefore rule out certain hot-, warm- or cold-start models \citep{Marleau_2014}.

PDS 70 is a young ($5.4 \pm 1.0$ Myr, \citealt{Mueller_2018}) system containing two directly-imaged planets, PDS 70 \textit{b} \citep{Keppler_2018} and PDS 70 \textit{c} \citep{Haffert_2019}, orbiting within a protoplanetary disk \citep{Riaud_2006}. While the age of this system would allow us to differentiate between hot- and cold-start formation, these planets present a particular challenge for dynamical mass fitting, since the currently available astrometry covers a relatively small fraction of their total orbital periods. \citet{Wang_2021} determined a $2\sigma$ upper mass limit for \textit{b} of 10 $M_{\rm Jup}$, but were unable to constrain the mass of \textit{c} within their uniform prior of 1--15 $M_{\rm Jup}$. Given its ability to achieve astrometric precisions down to 50~\mas \citep{GRAVITY2017, GRAVITY_2020}, the GRAVITY instrument on the Very Large Telescope Interferometer (VLTI) is well placed to provide the precise relative astrometry necessary for dynamical mass estimates in exoplanetary systems \citep[e.g.,][]{Balmer_2024}. 

Recently, \citet{Christiaens_2024} reported the detection of a point-like signal located near the outer edge of the PDS 70 inner disk compatible with the presence of a third protoplanet in the system. This new planet candidate, referred to from here as PDS 70 \textit{d}, is consistent with similar features observed with VLT/SPHERE high-contrast imaging by \citet{Mesa_2019} and re-reductions of VLT/SINFONI, VLT/NaCo and VLT/SPHERE imaging \citep{Hammond_2025}. However, current observations do not eliminate the possibility that this signal belongs to a dusty feature of the protoplanetary disc.

In this paper, we use new GRAVITY observations of PDS 70 \textit{b} and \textit{c} to fit orbital parameters to these planets, including updated dynamical mass estimates. The observations are described in Section~\ref{sec:observations}. In Section~\ref{sec:2planets} we fit orbital parameters to a two-planet system described by astrometry from \textit{b} and \textit{c} only. In Section~\ref{sec:3planets} we fit orbital parameters to a three-planet system described by astrometry from \textit{b}, \textit{c} and the candidate \textit{d}. In Section~\ref{sec:evo_models} we use our dynamical mass estimates from both scenarios to constrain $S_i$ for these planets, and determine which of the hot-, warm-, and cold-start evolutionary models are consistent with these masses. We discuss our results in Section~\ref{sec:discussion} and our conclusions are given in Section~\ref{sec:conclusions}.

\section{Observations and Data Reduction}
\label{sec:observations}

\subsection{GRAVITY Observations}

The new astrometric measurements presented in this paper were obtained between Jan 2021 and Feb 2022 with the GRAVITY instrument \citep{GRAVITY2017} and the four 8m Unit Telescopes (UTs) of ESO's VLTI. The star PDS 70~A itself served as the visual natural guide star for the adaptive optics and as the $K$-band fringe tracking reference source. We collected data using the $K$-band medium resolution mode ($R\approx500$). The observing set-up followed the strategy outlined in \citet{Nowak_2020} and \citet{Wang_2021}. Table~\ref{tab:obs_log} gives an overview on the observing dates, integration times on each planet, and atmospheric conditions. In our analysis (see Section \ref{sec:orbital_dynamics}) we combined this new data with GRAVITY astrometry from \citet{Wang_2021}.

\begin{table*}
\caption{Observing log of the new GRAVITY astrometric measurements reported in this paper. MJD corresponds to the halftime of each observing sequence.}
\begin{centering}
    \label{tab:obs_log}
    \begin{tabular}{lccccccr}
         \hline \hline
Date&MJD&  Planet & DIT/NDIT/NEXP& Airmass& Seeing & $\tau_0$ & Proposal ID \\
 & & & & & [arcsec] & [ms] &\\ \hline
2021-01-07\tablefootmark{a}&	59221.37&	\textit{b} &100s/4/3	&1.20--1.38 &0.70--1.21 &4.6--6.4 &	1104.C-0651\\
2021-01-08&	59222.34&	\textit{c} &100s/4/5	&1.26--1.69	&0.37--0.65 &4.8--8.6 &		1104.C-0651\\
2021-03-28&	59301.24&	\textit{c} &100s/4/4	&1.05--1.09	&0.66--0.94 &3.8--5.0 & 	105.209D\\
2021-03-29&	59302.25&	\textit{c} &100s/4/4	&1.04--1.07	&0.42--0.56 &5.4--7.1 & 	105.209D\\
2021-04-03&	59307.23&	\textit{c} &100s/4/4	&1.04--1.08	&0.48--0.76 &4.1--7.1 & 	105.209D\\
2021-05-29&	59363.10&	\textit{c} &100s/4/8	&1.04--1.08	&0.48--0.98 &3.3--6.3 & 	105.209D\\
2021-05-31&	59365.03&	\textit{c} &100s/4/4	&1.07--1.17	&0.59--0.79 &2.4--4.8 & 	105.209D\\
2022-02-21&	59631.28&	\textit{b} &100s/4/12&1.05--1.53	&0.57--1.41 &1.9--9.1 & 	1104.C-0651\\
2022-02-21&	59631.28&	\textit{c} &100s/4/7	&1.07--1.33	&0.61--1.23 &3.3--8.5 & 	1104.C-0651\\
\hline
\end{tabular}
 \tablefoot{
    \tablefoottext{a}{Data of lower quality. This epoch has been ignored in the subsequent analysis.}
}
\end{centering}
\end{table*}

\subsection{Reduction of Relative Astrometry}

The data were reduced using the standard GRAVITY pipeline \citep{GRAVITY_DRS2014} resulting in ``astroreduced'' OIFITS (Optical Interferometry FITS) files. Subsequent analysis used the Python scripts \verb<cleanGRAVITY< and \verb<exoGRAVITY< \citep{Nowak_2020} to deduce the relative astrometry between the planet and the star. The resulting measurements are reported in Table~\ref{tab:astro_new}. Typical uncertainties in the relative position of PDS\,70\,\textit{c} amount to a few 100~\mas. For PDS\,70\,\textit{b}, the analysis of the 2021-01-07 data set resulted in uncertainties on the combined relative astrometry larger than 6\,mas. We attribute this to a combination of relatively short total integration time, varying and unfavourable observing conditions, and the challenge of observing a planet separated by less than three times the single UT $K$-band diffraction limit from its host star. Given the large uncertainty, we ignored this epoch in the subsequent analysis. The 2022-02-21 data set on PDS\,70\,\textit{b} constitutes a four times longer observing sequence, with 2/3 of the data obtained in good to very good observing conditions, resulting in relative astrometry with uncertainties of $\lessapprox$1\,mas. All of the astrometry for PDS 70 \textit{c} was obtained in good observing conditions and so none of it needed to be discarded.

\setlength{\tabcolsep}{5pt}
\begin{table} 
  \caption{Astrometry of PDS 70 b and c relative to the central star.} 
  \centering 
  \begin{tabular}{c c c c c c c} 
    \hline 
    & MJD & $\Delta{\rm R.A.}$ & $\sigma_{\Delta{\rm R.A.}}$ & $\Delta{\rm Decl.}$ &  $\sigma_{\Delta{\rm Decl.}}$ & $\rho$ \\
    & (days) & (mas) & (mas) & (mas) & (mas) \\
    \hline 
    \textit{b} & 59631.28 & 111.14 & 1.0 & $-115.78$ & 0.83 & -0.58 \\
    \hline
    \textit{c} & 59222.34 & $-212.88$ & 0.52 & 19.32 & 0.24 & 0.25 \\ 
    \textit{c} & 59301.24 & $-212.88$ & 0.08 & 17.85 & 0.14 & -0.20 \\ 
    \textit{c} & 59302.25 & $-212.83$ & 0.13 & 17.67 & 0.10 & -0.97 \\ 
    \textit{c} & 59307.23 & $-212.88$ & 0.16 & 17.55 & 0.23 & -0.91 \\ 
    \textit{c} & 59363.10 & $-212.58$ & 0.15 & 16.27 & 0.15 & -0.69 \\ 
    \textit{c} & 59365.03 & $-212.32$ & 0.13 & 16.15 & 0.11 & -0.21 \\ 
    \textit{c} & 59631.28 & $-211.75$ & 0.53 & 10.04 & 0.37 & -0.85 \\ 
    \hline 
  \end{tabular}
   \tablefoot{
    \tablefoottext{}{A correlation coefficient ($\rho$) is necessary to describe the alignment of the uncertainties of these interferometric observations with the on-sky coordinates. The covariance matrix can be reconstructed with $\sigma_{\Delta{\rm R.A.}}^2$ and $\sigma_{\Delta{\rm Decl.}}^2$ on the diagonal, and $\rho\sigma_{\Delta{\rm R.A.}}\sigma_{\Delta{\rm Decl.}}$ on the off-diagonal.}
}
  \label{tab:astro_new} 
\end{table} 

\section{Orbital Dynamics}
\label{sec:orbital_dynamics}

\subsection{Orbital Fitting with MCMC}
\label{sec:2planets}

We fit orbital parameters to PDS 70 \textit{b} and \textit{c} with the \verb|orbitize| package \citep[version 3.0.0;][]{Blunt_2020} using a combination of our new GRAVITY astrometry, astrometry from a new reduction of VLT/SINFONI data (described in Appendix~\ref{sec:SINFONI}), and literature astrometry from Gemini/NICI \citep{Keppler_2018}, VLT/NaCo \citep{Keppler_2018}, VLT/SPHERE/IRDIS \citep{Keppler_2018, Mueller_2018, Mesa_2019, Wahhaj2024}, VLT/SINFONI \citep{Christiaens2019}, VLT/MUSE \citep{Haffert_2019}, Keck/NIRC2 \citep{Wang_2020}, VLT/GRAVITY \citep{Wang_2021}, JWST/NIRISS \citep{Blakely_2024}, JWST/NIRCam \citep{Christiaens_2024} and MagAO-X \citep{Close2025}. 
For a complete list of the astrometry used from the literature, see Table~\ref{tab:bcd_lit_astrometry} in Appendix~\ref{sec:lit_data}. 

\verb|orbitize!| uses the parallel-tempered affine-invariant Markov chain Monte Carlo (MCMC) sampler \verb|ptemcee| \citep{Foreman-Mackey_2013, Vousden_2016} for orbital fitting. Parallel-tempered MCMC utilizes multiple Markov chains at different "temperatures". With increasing temperature, the posterior distribution asymptotes to the prior, preventing the sampler from becoming stuck within well separated modes. We used 20 temperatures, with 1000 walkers at each temperature and 20,000 steps per walker. Convergence was assessed by visual inspection of the walker chains. In order to remove burn-in, the last 10,000 steps from each walker (10 million parameter sets in total) were used to form the posterior probability distribution. 

We fit the following orbital parameters for each planet: semimajor axis ($a$), eccentricity ($e$), inclination ($i$), argument of periastron ($\omega$), longitude of the ascending node ($\Omega$), epoch of periastron in units of the fractional orbital period ($\tau$), and mass ($M$). Inclinations above 90 degrees refer to clockwise orbits, while inclinations below 90 degrees refer to counter-clockwise orbits. MJD 58,849 (2020-01-01) was used as the reference epoch for $\tau$ for both planets. Subscripts on the parameters indicate whether they belong to PDS 70 \textit{b} or \textit{c}. We also fit the stellar mass ($M_*$) and system parallax.

The priors used for each parameter are listed in Table~\ref{tab:2body_params}. The priors for all the planetary parameters, with the exception of mass, match those from \citet{Wang_2020} and \citet{Wang_2021} and are designed to be uninformative. Given that the 95\% confidence interval (CI) on the planetary masses from \citet{Wang_2021} covers the majority of their $1 - 15$ $M_{\rm Jup}$ mass uniform prior, we extend the upper edge of this prior to $30$ $M_{\rm Jup}$. We also set the lower end of the prior to zero, since planets less massive than $1$ $M_{\rm Jup}$ can still open gaps in protoplanetary disks \citep{Rosetti_2016}. We set a Gaussian prior for parallax based on the system parallax of $8.8975 \pm 0.0191$ mas from Gaia DR3 \citep{Gaia_2021}. We also used the same Gaussian prior for the stellar mass as \citet{Wang_2021}, with a mean of $0.88$ $M_\odot$ and a standard deviation of $0.09$ $M_\odot$, based on their stellar SED fit with 10\% errors to account for systematics of their model.

\begin{table*}
    \caption{Orbital parameters fit to astrometry of PDS 70 \textit{b} and \textit{c} only. The third column shows the results of the fit using the Gaussian prior on the mutual inclination (see Equation~\ref{eq:coplanarprior}), while the final two columns show the results of the fit using Equation~\ref{eq:stabprior} as a prior, before and after the $N$-body analysis. For each parameter the median value of the posterior is listed, with subscripts and superscripts listing the 68\% and 95\% CIs (with the 95\% CIs in parentheses).}
    \label{tab:2body_params}
    \centering
    \begin{tabular}{c|c|c|c|c}
         \hline
         \hline
         Quantity & Prior & Coplanar & Stable & Stable (incl. $N$-body) \\
         \hline
         $a_b$ (au) & LogUniform(1, 100) & $20.1^{+0.7 (+1.5)}_{-0.7 (-1.4)}$ & $20.8^{+0.6 (+1.3)}_{-0.7 (-1.4)}$ & $20.7^{+0.6 (+1.4)}_{-0.5 (-1.2)}$ \\
         $e_b$ & Uniform(0, 1) & $0.21^{+0.05 (+0.10)}_{-0.05 (-0.10)}$ & $0.18^{+0.05 (+0.12)}_{-0.05 (-0.09)}$ & $0.16^{+0.05 (+0.12)}_{-0.04 (-0.01)}$ \\
         $i_b$ (deg) & $\sin(i)$ & $130.7^{+1.9 (+3.8)}_{-1.9 (-3.7)}$ & $130.3^{+1.8 (+3.9)}_{-1.8 (-3.4)}$ & $130.6^{+1.6 (+3.5)}_{-1.6 (-3.2)}$ \\
         $\omega_b$ (deg) & Uniform(0, 360) & $197^{+8 (+14)}_{-8 (-19)}$ & $193^{+9 (+16)}_{-10 (-22)}$ & $190^{+10 (+16)}_{-13 (-28)}$ \\
         $\Omega_b$ (deg) & Uniform(0, 360) & $178.2^{+4.5 (+9.1)}_{-4.1 (-8.0)}$ & $179^{+5 (+11)}_{-4 (-8)}$ & $176^{+5 (+12)}_{-4 (-8)}$ \\
         $\tau_b$ & Uniform(0, 1) & $0.373^{+0.024 (+0.047)}_{-0.021 (-0.040)}$ & $0.360^{+0.027 (+0.051)}_{-0.023 (-0.045)}$ & $0.355^{+0.028 (+0.055)}_{-0.025 (-0.049)}$ \\
         \hline
         $a_c$ (au) & LogUniform(1, 100) & $31.9^{+0.9 (+1.8)}_{-1.0 (-2.0)}$ & $33.3^{+1.1 (+2.0)}_{-1.0 (-2.2)}$ & $33.9^{+1.0 (+1.7)}_{-1.1 (-2.1)}$ \\
         $e_c$ & Uniform(0, 1) & $0.065^{+0.017 (+0.036)}_{-0.015 (-0.029)}$ & $0.046^{+0.014 (+0.032)}_{-0.013 (-0.027)}$ & $0.042^{+0.014 (+0.027)}_{-0.013 (-0.030)}$ \\
         $i_c$ (deg) & $\sin(i)$ & $131.5^{+1.2 (+2.5)}_{-1.1 (-2.2)}$ & $130.4^{+1.3 (+2.8)}_{-1.2 (-2.2)}$ & $129.8^{+1.3 (+2.7)}_{-1.0 (-1.9)}$ \\
         $\omega_c$ (deg) & Uniform(0, 360) & $89^{+9 (+21)}_{-10 (-24)}$ & $77^{+11 (+33)}_{-11 (-22)}$ & $77^{+12 (+28)}_{-13 (-25)}$ \\
         $\Omega_c$ (deg) & Uniform(0, 360) & $154.9^{+2.1 (+4.0)}_{-2.6 (-5.8)}$ & $157.5^{+1.8 (+3.3)}_{-2.2 (-5.8)}$ & $158.0^{+1.7 (+3.1)}_{-2.0 (-3.8)}$ \\
         $\tau_c$ & Uniform(0, 1) & $0.572^{+0.030 (+0.068)}_{-0.035 (-0.078)}$ & $0.52^{+0.04 (+0.11)}_{-0.03 (-0.07)}$ & $0.521^{+0.038 (+0.085)}_{-0.038 (-0.070)}$ \\
         \hline
         Parallax (mas) & $\mathcal{N}$(8.8975, 0.0191) & $8.880^{+0.024 (+0.048)}_{-0.024 (-0.051)}$ & $8.881^{+0.025 (+0.048)}_{-0.023 (-0.046)}$ & $8.878^{+0.026 (+0.050)}_{-0.023 (-0.046)}$ \\
         $M_b$ ($M_{\rm Jup}$) & Uniform(0, 30) & $10^{+10 (+18)}_{-7 (-10)}$ & $6^{+6 (+18)}_{-4 (-6)}$ & $1.4^{+2.1 (+4.2)}_{-1.0 (-1.3)}$ \\
         $M_c$ ($M_{\rm Jup}$) & Uniform(0, 30) & $9^{+8 (+18)}_{-6 (-9)}$ & $9^{+9 (+19)}_{-6 (-8)}$ & $6.4^{+4.3 (+8.8)}_{-3.7 (-5.7)}$ \\
         $M_*$ ($M_\odot$) & $\mathcal{N}$(0.88, 0.09) & $0.967^{+0.033 (+0.067)}_{-0.030 (-0.060)}$ & $0.954^{+0.032 (+0.070)}_{-0.031 (-0.063)}$ & $0.952^{+0.032 (+0.065)}_{-0.028 (-0.063)}$ \\
         \hline
    \end{tabular}
\end{table*}

\subsubsection{Coplanarity vs Stability}
\label{sec:coplanar_stable}

In order to constrain the orbital parameter space for the system, we produced two orbital fits, one assuming stability of the orbits and the other assuming coplanarity between the orbits and the protoplanetary disk. In the case where we assumed stability, we imposed the prior given by Equation(1) of \citet{Wang_2020}, ensuring the non-crossing of the orbits of \textit{b} and \textit{c}:
\begin{equation}
 \label{eq:stabprior}
    a_c (1-e_c) > a_b (1+e_b).
\end{equation}

\begin{figure*}
    \centering
    \includegraphics[width=0.9\linewidth]{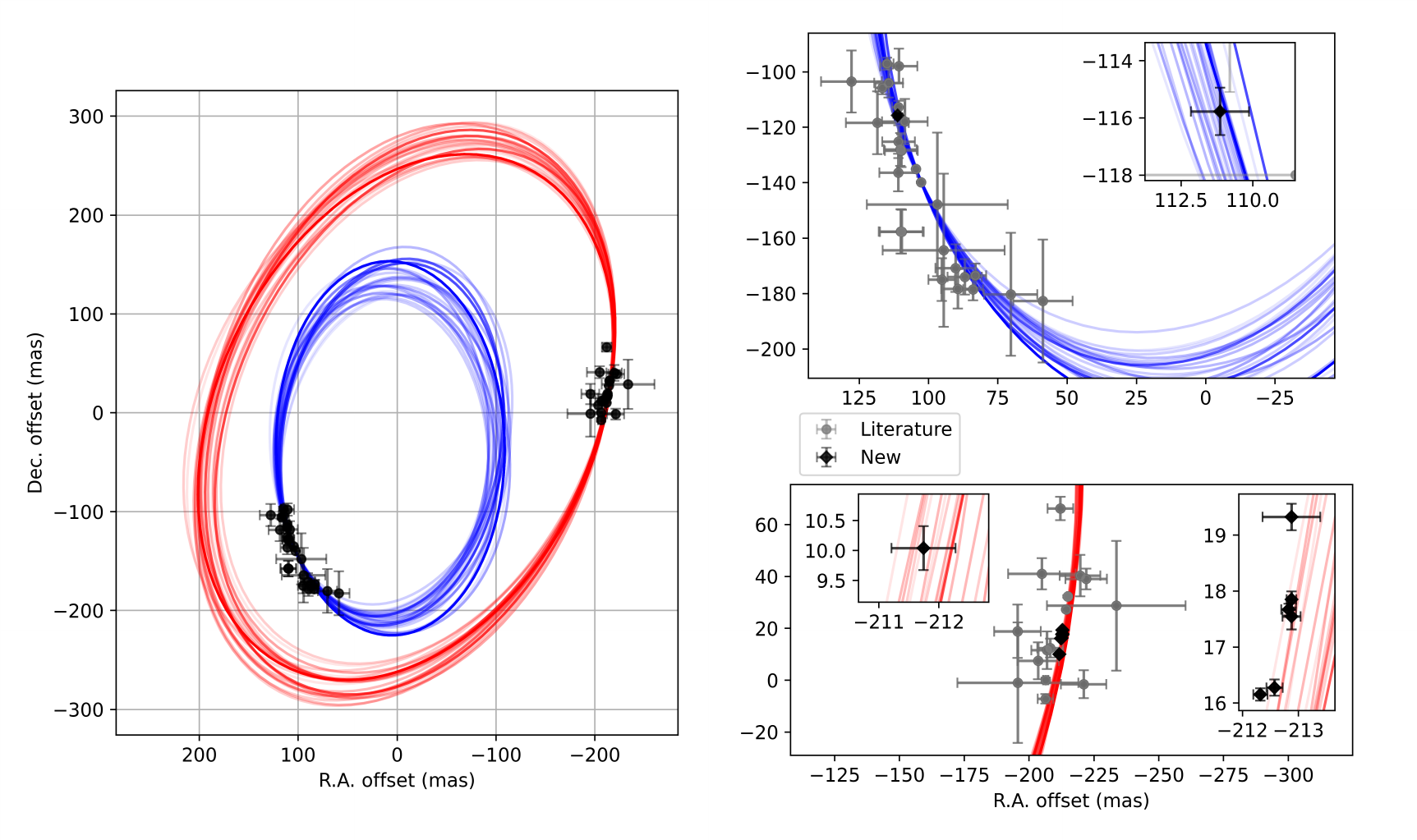}
    \caption{Sample orbits from the posterior of the coplanar fit to PDS 70 \textit{b} and \textit{c} only, plotted alongside the astrometry used in the fit. Sample orbits for \textit{b} are shown in blue, while sample orbits for \textit{c} are shown in red. Zoomed in portions of the orbit, containing all the astrometry used in the fit, are shown in the right hand panels. The new GRAVITY astrometry (see Table~\ref{tab:astro_new}) is shown in black and literature astrometry is show in gray. The inset panels zoom-in on the new astrometry. R.A. and Dec. are given relative to the central star}.
    \label{fig:2planet_orbits}
\end{figure*}

In the scenario assuming coplanarity we placed a constraint on the mutual inclination between the two planets, as well as each planet and the PDS 70 protoplanetary disc. The mutual inclination $\Phi_{12}$ between two orbital planes (with inclinations $i_1$ and $i_2$ and longitudes of the ascending node $\Omega_1$ and $\Omega_2$) is given by the law of cosines (e.g.,  \citealp{Bean_2009}):
\begin{equation}
    \label{eq:coplanarprior}
    \cos (\Phi_{12}) = \cos(i_1)\cos(i_2) + \sin(i_1)\sin(i_2)\cos(\Omega_1 - \Omega_2).
\end{equation}
Figure~\ref{fig:2planet_orbits} shows sample orbits taken from the posterior of the coplanar fit, plotted alongside the astrometry used for the fit.

We set the inclination and position angle (PA) of the disk to $i=128.3$ degrees (equivalent to $i=51.7$ degrees but for clockwise orbits) and $PA=156.7$ degrees based on the values from \citet{Keppler_2019}. In the coplanar scenario, we placed Gaussian priors on all three of the mutual inclinations between \textit{b}, \textit{c}, and the protoplanetary disc, with a mean of 0 degrees and a standard deviation of 10 degrees, matching the prior on the coplanar model used in \citet{Wang_2020}.

In order to reduce convergence time, for both the stable and coplanar scenarios, we set the initial positions of the MCMC walkers using random draws from a Gaussian centred at the respective median parameters found by \citet{Wang_2021} and with standard deviations matching the size of the largest $68\%$ CI for each parameter.

For both the stable and coplanar fits, the median of the posterior for each parameter, as well as the 68\% and 95\% CIs on either side are shown in Table~\ref{tab:2body_params}. The full posteriors for the orbital parameters of \textit{b} and \textit{c} are shown in Appendix \ref{sec:corner_plots} in Figures \ref{fig:corner_bc_b} and \ref{fig:corner_bc_c} respectively. In all cases the inclination of the disk falls within the 95\% CIs for both \textit{b} and \textit{c}, indicating that both planets are coplanar with the disk.

\subsubsection[N-body Stability Analysis]{\textit{N}-body Stability Analysis}
\label{sec:nbody}

While the fulfillment of Equation~\ref{eq:stabprior} is necessary for the stability of the PDS 70 planets, it does not ensure long-term stability. Therefore we performed an additional step of $N$-body stability analysis on the parameter sets resulting from the MCMC fit where this prior was used. We used the \verb|REBOUND| $N$-body package \citep{rebound} with IAS15, a 15th order Gauss--Radau integrator \citep{reboundias15}, to evolve these parameter sets backwards by 2.0~Myr. Parameter sets were deemed to be unstable if
\begin{enumerate}[label=\alph*)]
    \item the planets passed within one Hill radius ($\sim 3.3$~au) of each other or
    \item any of the planets were ejected to more than 40~au from their host star.
\end{enumerate}
We chose 40~au as the ejection radius since it is the location of the outer edge of the dust gap in the PDS 70 system.

We chose a timescale of 2.0~Myr to match the lower bound of the type II migration timescale for the PDS 70 planets. The migration timescale is the time required for a planet to migrate inwards all the way to its host star due to gas drag from the circumstellar disk \citep{Ida_2018}. A planet still embedded in a gas disk undergoes type I migration \citep{Tanaka_2002}, whose timescale (neglecting contributions other than the Lindblad torque) is given by
\begin{equation}
    t_{\rm mig1} \simeq \frac{1}{2c} \left(\frac{M_p}{M_*}\right)^{-1} \left(\frac{\Sigma r^2}{M_*}\right)^{-1} \left(\frac{H}{r}\right)^2 \Omega^{-1},
\end{equation}
where $c \sim 2$ \citep{Kanagawa_2018} is a constant, $M_p$ is the planet mass, $\Sigma$ is the  surface density of the circumstellar disk, $r$ is the distance from the star, $H$ is the gas scale height of the disk and $\Omega$ is the Keplerian frequency. A planet inside a disk gap undergoes type II migration \citep{Kanagawa_2018}, whose timescale is given by
\begin{equation}
    t_{\rm mig2} \simeq \frac{\Sigma}{\Sigma_{\rm min}} t_{\rm mig1},
\end{equation}
where $\Sigma_{\rm min}$ is the gas surface density in the gap. We used the radial parameterizations of $\Sigma$ and $h$, as well as $\Sigma_{\rm min}$, from \citet{Portilla-Revelo_2023} to determine that $t_{\rm mig2}~\gtrapprox~2.0$~Myr for the PDS 70 planets.

Of the 10 million available parameter sets from the results of the MCMC fit assuming non-crossing orbits $\sim 8.9$ million were determined to be unstable as a result of the $N$-body stability analysis. The final column of Table~\ref{tab:2body_params} shows the median for each parameter after this stability analysis, as well as the 68\% and 95\% CIs on each side.

\begin{figure*}
    \centering
    \includegraphics[width=0.9\linewidth]{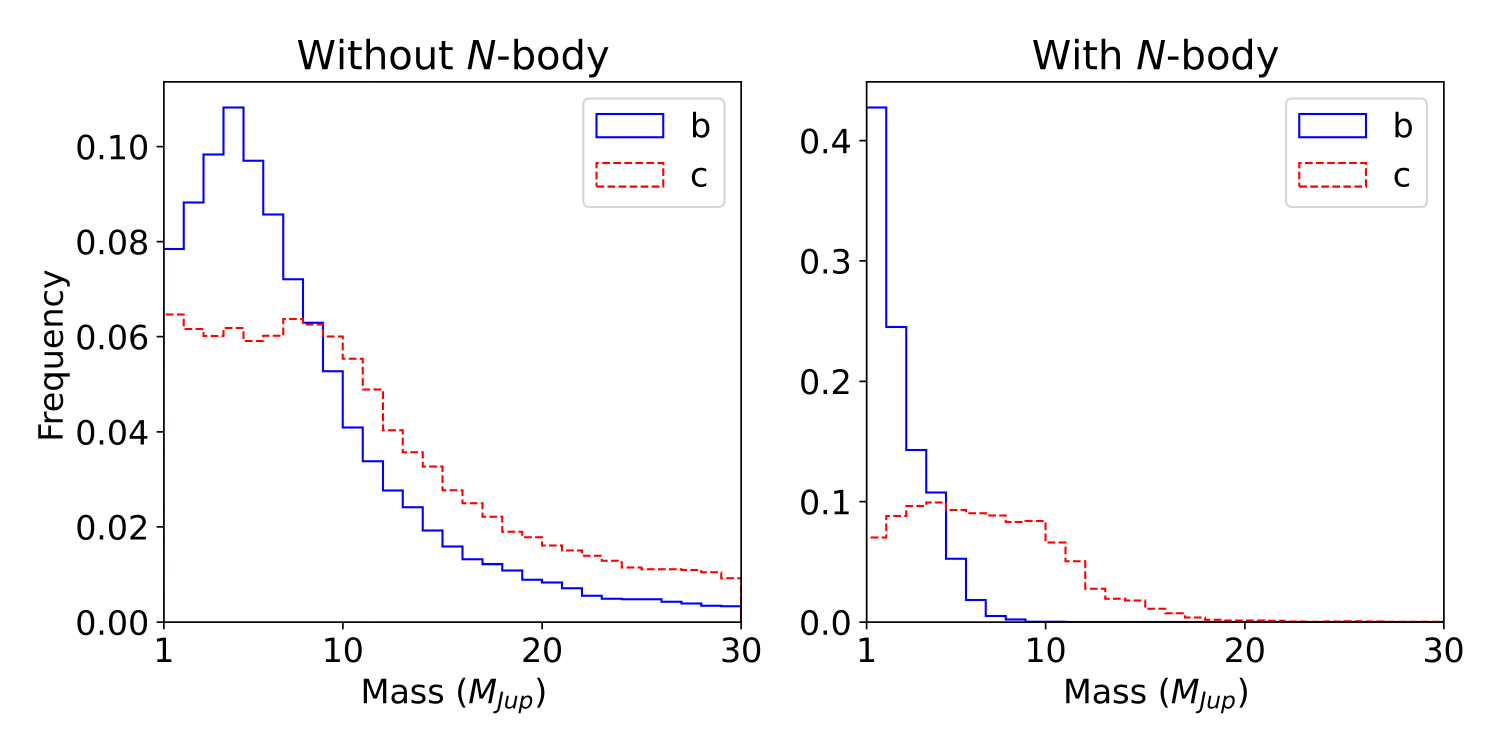}
    \caption{Posterior distributions of the planet masses from the stable fit to PDS 70 \textit{b} and \textit{c} only, before and after $N$-body stability analysis. The mass posteriors for \textit{b} are shown in blue (solid line), while the posteriors for \textit{c} are shown in red (dashed line).}
    \label{fig:2planet_masses}
\end{figure*}

Figure~\ref{fig:2planet_masses} shows the posterior distributions of the planetary masses, for the fit assuming stability, before and after performing $N$-body stability analysis. Using these parameter sets, we derived 95\% confidence upper mass limits of $4.9$ $M_{\rm Jup}$ for PDS 70 \textit{b} and $13.6$ $M_{\rm Jup}$ for PDS 70 \textit{c}. These mass limits are defined such that only the highest 5\% of the mass posterior lies above them. Without the $N$-body analysis, the posteriors of both the coplanar and stable fits are not informative enough to constrain the planet masses, since the 95\% CIs cover most of the prior.

\subsection{Two Planets or Three?}
\label{sec:3planets}

\citet{Christiaens_2024} suggest the existence of a third protoplanet candidate, PDS 70 \textit{d} in the PDS 70 system, from the detection of a point-like signal near the outer edge of the inner disk. While this signal is consistent with similar features observed by \citet{Mesa_2019} and \citet{Hammond_2025}, we cannot eliminate the possibility that it belongs to a disk feature. In this section we fit orbits to \textit{b}, \textit{c} and the candidate \textit{d} to assess possible orbital configurations and their stability.

We used the same orbital fitting procedure and MCMC settings as described in Section~\ref{sec:2planets} to fit orbits to a three planet system described by astrometry of PDS 70 \textit{b}, \textit{c} and the candidate \textit{d}. In addition to the astrometry described in Section~\ref{sec:2planets}, we used astrometry from \citet{Christiaens_2024} and \citet{Hammond_2025} to fit the orbit of \textit{d} (listed in Table~\ref{tab:bcd_lit_astrometry} in Appendix~\ref{sec:lit_data}).

In addition to the orbital parameters described in Section~\ref{sec:2planets}, we also fit $a$, $e$, $i$, $\omega$, $\Omega$, $\tau$ and $M$ to \textit{d}, with priors matching those for \textit{b} and \textit{c}. In the case of the parameters shared with the two-planet model, we set the initial positions of the MCMC walkers as described in Section~\ref{sec:2planets}. In the case of the parameters specific to PDS 70 \textit{d}, we set the initial positions of the walkers by fitting these parameters to the PDS 70 \textit{d} astrometry alone using \verb|orbitize!|, with the same MCMC setting as described in Section~\ref{sec:2planets}. The positions of the walkers for the fit to \textit{b}, \textit{c} and the candidate \textit{d} were set using random draws from a Gaussian centered at the median values found by this fit, with a standard deviation equivalent to the size of the largest 68\% CI for each parameter.

We produced two orbital fits using this astrometry, one assuming coplanarity and one assuming stability, with conditions identical to those described in Section~\ref{sec:coplanar_stable}. The same type of $N$-body stability analysis as described in Section~\ref{sec:nbody} was applied to the fit assuming stability, in order to further constrain the parameters. Out of the 10 million available parameter sets from the MCMC orbital fit, only $\sim 6000$ were determined to be stable orbital configurations.

\begin{figure*}
    \centering
    \includegraphics[width=0.9\linewidth]{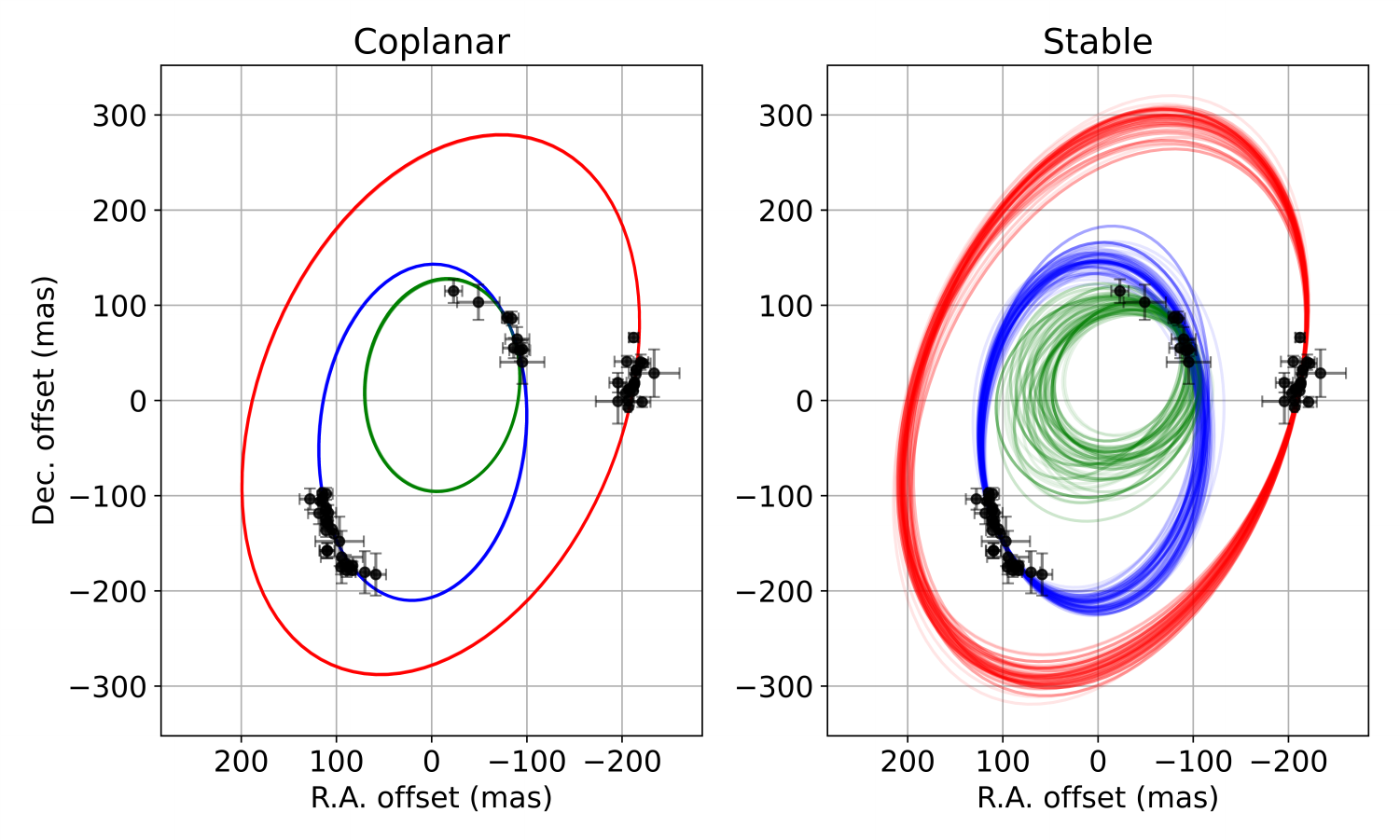}
    \caption{Sample orbits from the fits to astrometry of PDS 70 \textit{b}, \textit{c} and the candidate \textit{d}. The left panel shows orbits taken from the fit assuming coplanarity, while the right panel shows orbits taken from the fit assuming stability (before $N$-body analysis). Orbits for \textit{b}, \textit{c} and the candidate \textit{d} are shown in blue, red and green respectively. The astrometry data is plotted with error bars in black. All astrometry is given relative to the central star. The coplanar fit produces many orbital configurations where the orbits of \textit{b} and the candidate \textit{d} overlap, even though these orbits would likely be unstable.}
    \label{fig:3planet_orbits}
\end{figure*}

Table~\ref{tab:3body_params} shows the median values of each parameter, as well as the 68\% and 95\% CIs, for both the coplanar and stable fits. Full posteriors for the orbital parameters of \textit{b}, \textit{c} and the candidate \textit{d} are shown in Appendix \ref{sec:corner_plots} in Figures \ref{fig:corner_bcd_b},\ref{fig:corner_bcd_c} and \ref{fig:corner_bcd_d} respectively. Sample orbits taken from both the coplanar and stable fits are shown in Figure~\ref{fig:3planet_orbits}.

\begin{table*}
    \caption{Same as Table~\ref{tab:2body_params}, but with orbital parameters fit to astrometry of PDS 70 \textit{b}, \textit{c} and the candidate \textit{d}.}
    \label{tab:3body_params}
    \centering
    \begin{tabular}{c|c|c|c|c}
         \hline
         \hline
         Quantity & Prior & Coplanar & Stable & Stable (incl. $N$-body) \\
         \hline
         $a_b$ (au) & LogUniform(1, 100) & $19.8^{+0.6 (+1.1)}_{-0.6 (-1.1)}$ & $20.6^{+0.5 (+1.1)}_{-0.5 (-1.1)}$ & $21.1^{+0.4 (+0.6)}_{-0.8 (-1.0)}$ \\
         $e_b$ & Uniform(0, 1) & $0.207^{+0.041 (+0.085)}_{-0.041 (-0.080)}$ & $0.183^{+0.05 (+0.09)}_{-0.04 (-0.09)}$ & $0.131^{+0.035 (+0.069)}_{-0.037 (-0.048)}$ \\
         $i_b$ (deg) & $\sin(i)$ & $131.7^{+1.9 (+3.9)}_{-1.8 (-3.4)}$ & $130.8^{+1.7 (+3.5)}_{-1.7 (-3.4)}$ & $128.7^{+1.9 (+3.4)}_{-1.2 (-3.7)}$ \\
         $\omega_b$ (deg) & Uniform(0, 360) & $191^{+6 (+16)}_{-5 (-10)}$ & $194^{+5 (+11)}_{-6 (-15)}$ & $191.4^{+2.7 (+4.9)}_{-3.6 (-5.1)}$ \\
         $\Omega_b$ (deg) & Uniform(0, 360) & $175.7^{+4.4 (+9.6)}_{-3.3 (-6.1)}$ & $177.9^{+4.4 (+9.2)}_{-3.7 (-7.6)}$ & $174.3^{+3.4 (+8.1)}_{-2.9 (-3.8)}$ \\
         $\tau_b$ & Uniform(0, 1) & $0.366^{+0.021 (+0.041)}_{-0.018 (-0.034)}$ & $0.365^{+0.022 (+0.043)}_{-0.017 (-0.034)}$ & $0.387^{+0.008 (+0.010)}_{-0.024 (-0.039)}$ \\
         \hline
         $a_c$ (au) & LogUniform(1, 100) & $32.5^{+0.9 (+1.8)}_{-0.9 (-1.9)}$ & $34.6^{+0.9 (+1.6)}_{-1.0 (-2.3)}$ & $35.3^{+0.8 (+1.0)}_{-0.9 (-2.2)}$ \\
         $e_c$ & Uniform(0, 1) & $0.062^{+0.019 (+0.039)}_{-0.014 (-0.028)}$ & $0.036^{+0.014 (+0.028)}_{-0.010 (-0.021)}$ & $0.033^{+0.011 (+0.016)}_{-0.008 (-0.017)}$ \\
         $i_c$ (deg) & $\sin(i)$ & $130.5^{+1.2 (+2.8)}_{-1.1 (-2.0)}$ & $129.1^{+1.2 (+2.8)}_{-1.0 (-1.8)}$ & $128.5^{+1.4 (+2.9)}_{-0.7 (-0.8)}$ \\
         $\omega_c$ (deg) & Uniform(0, 360) & $94^{+8 (+14)}_{-18 (-30)}$ & $66^{+12 (+38)}_{-10 (-21)}$ & $63^{+19 (+23)}_{-13 (-21)}$ \\
         $\Omega_c$ (deg) & Uniform(0, 360) & $155.3^{+2.0 (+3.6)}_{-2.3 (-5.0)}$ & $158.8^{+1.6 (+3.3)}_{-2.3 (-6.0)}$ & $159.8^{+1.0 (+2.2)}_{-0.8 (-2.9)}$ \\
         $\tau_c$ & Uniform(0, 1) & $0.585^{+0.026 (+0.047)}_{-0.055 (-0.097)}$ & $0.49^{+0.04 (+0.13)}_{-0.03 (-0.06)}$ & $0.482^{+0.049 (+0.063)}_{-0.051 (-0.066)}$ \\
         \hline
         $a_d$ (au) & LogUniform(1, 100) & $12.9^{+1.6 (+3.5)}_{-1.6 (-2.9)}$ & $10.7^{+1.5 (+2.6)}_{-1.3 (-2.3)}$ & $10.7^{+2.1 (+2.2)}_{-1.1 (-2.3)}$ \\
         $e_d$ & Uniform(0, 1) & $0.21^{+0.12 (+0.24)}_{-0.10 (-0.18)}$ & $0.27^{+0.15 (+0.29)}_{-0.15 (-0.25)}$ & $0.25^{+0.13 (+0.30)}_{-0.23 (-0.25)}$ \\
         $i_d$ (deg) & $\sin(i)$ & $136^{+5 (+11)}_{-5 (-10)}$ & $153^{+13 (+22)}_{-12 (-21)}$ & $151^{+6 (+18)}_{-8 (-23)}$ \\
         $\omega_d$ (deg) & Uniform(0, 360) & $80^{+150 (+220)}_{-30 (-60)}$ & $80^{+110 (+220)}_{-60 (-80)}$ & $29^{+24 (+51)}_{-17 (-26)}$ \\
         $\Omega_d$ (deg) & Uniform(0, 360) & $161^{+12 (+25)}_{-18 (-37)}$ & $190^{+130 (+160)}_{-50 (-130)}$ & $144^{+21 (+38)}_{-10 (-22)}$ \\
         $\tau_d$ & Uniform(0, 1) & $0.58^{+0.12 (+0.21)}_{-0.40 (-0.54)}$ & $0.47^{+0.09 (+0.18)}_{-0.09 (-0.20)}$ & $0.55^{+0.04 (+0.09)}_{-0.08 (-0.15)}$ \\
         \hline
         Parallax (mas) & $\mathcal{N}$(8.8975, 0.0191) & $8.872^{+0.025 (+0.048)}_{-0.031 (-0.058)}$ & $8.881^{+0.024 (+0.045)}_{-0.025 (-0.058)}$ & $8.882^{+0.018 (+0.059)}_{-0.021 (-0.025)}$ \\
         $M_b$ ($M_{\rm Jup}$) & Uniform(0, 30) & $10^{+9 (+18)}_{-5 (-9)}$ & $5^{+4 (+11)}_{-3 (-4)}$ & $0.7^{+1.7 (+4.6)}_{-0.5 (-0.7)}$ \\
         $M_c$ ($M_{\rm Jup}$) & Uniform(0, 30) & $9^{+7 (+18)}_{-5 (-8)}$ & $8^{+8 (+18)}_{-5 (-7)}$ & $2.4^{+4.3 (+5.8)}_{-1.3 (-2.2)}$ \\
         $M_d$ ($M_{\rm Jup}$) & Uniform(0, 30) & $8^{+7 (+16)}_{-6 (-8)}$ & $9^{+9 (+18)}_{-7 (-9)}$ & $0.4^{+1.5 (+1.9)}_{-0.3 (-0.4)}$ \\
         $M_*$ ($M_\odot$) & $\mathcal{N}$(0.88, 0.09) & $0.960^{+0.032 (+0.065)}_{-0.031 (-0.063)}$ & $0.953^{+0.031 (+0.062)}_{-0.033 (-0.064)}$ & $0.965^{+0.041 (+0.068)}_{-0.025 (-0.088)}$ \\
         \hline
    \end{tabular}
\end{table*}

In all cases the inclination of the disk falls within the 95\% CIs for the inclination of \textit{b} and \textit{c}, again indicating that these planets are coplanar with the disk. This is also true for the candidate \textit{d} in the coplanar case, however both the stable and $N$-body case suggest somewhat higher inclinations for \textit{d}.

\begin{figure*}
    \centering
    \includegraphics[width=0.9\linewidth]{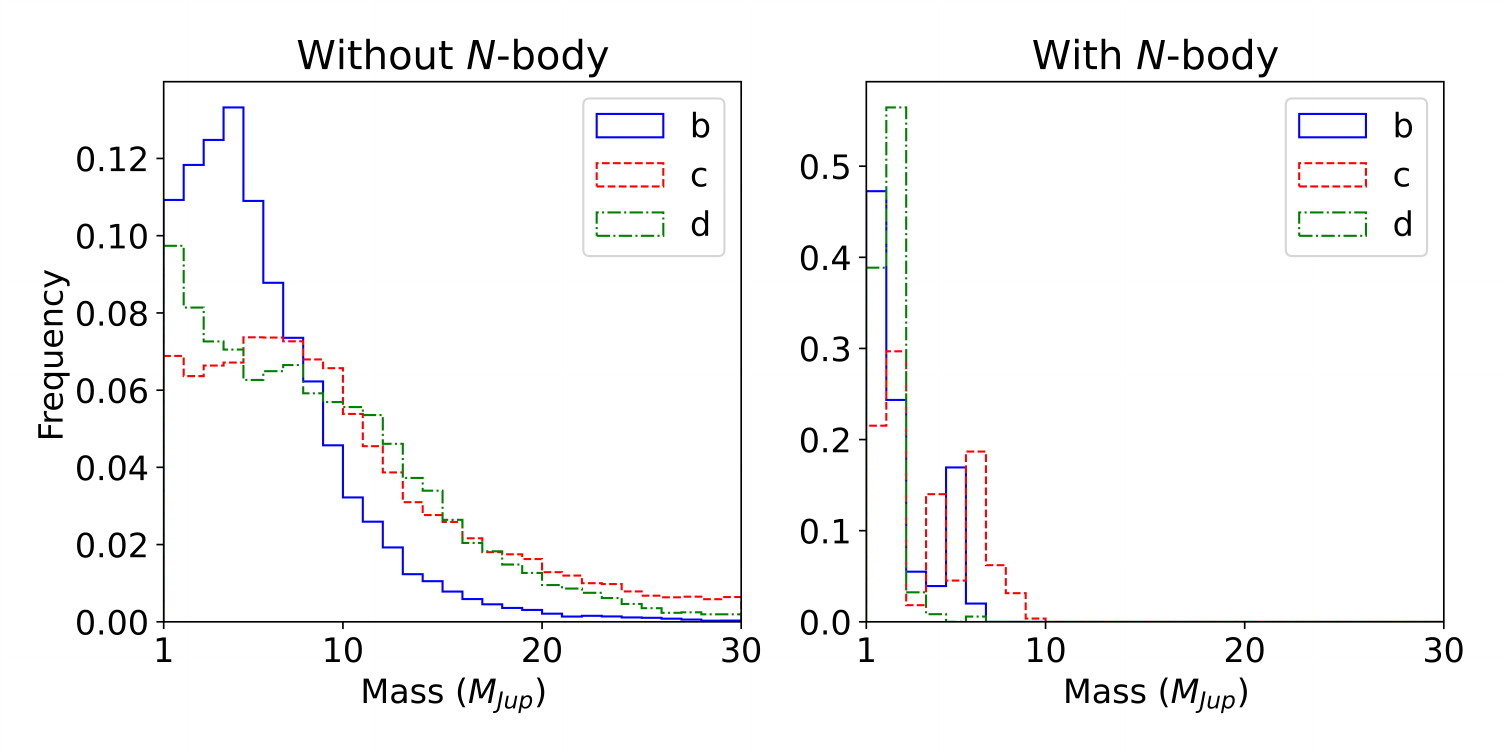}
    \caption{Posterior distributions of the planet masses from the stable fit to PDS 70 \textit{b}, \textit{c} and the candidate \textit{d}, before and after $N$-body stability analysis. The mass posteriors for \textit{b} are shown in blue (solid line), the posteriors for \textit{c} are shown in red (dashed line), and the posteriors for \textit{d} are shown in green (line with dots and dashes). This analysis assumes that \textit{d} is a planet and not a disk feature.}
    \label{fig:3planet_masses}
\end{figure*}

In this scenario the $N$-body stability analysis allows us to place new 95\% confidence upper mass limits for \textit{b} and \textit{c}. For \textit{b} we find an upper mass limit of 5.3 $M_{\rm Jup}$, while for \textit{c} we find an upper mass limit of 7.5 $M_{\rm Jup}$. In addition, for \textit{d} we find a 95\% confidence upper mass limit of 2.2 $M_{\rm Jup}$. Figure~\ref{fig:3planet_masses} shows the posterior distributions of the planetary masses both before and after performing $N$-body stability analysis.

\subsection{Period Ratios}

\citet{Wang_2021} give the period ratio of \textit{c} to \textit{b} as $2.13^{+0.27}_{-0.24}$ for the 68\% CI and $2.13^{+0.56}_{-0.45}$ for the 95\% CI, putting them near the 2:1 mean-motion resonance (MMR) and excluding to a few $\sigma$ other first-order MMRs such as 3:2 and 4:3. Table~\ref{tab:periods} lists the period ratios between each pair of planets that we found in our fits. Our fits to \textit{b} and \textit{c} only also find that the 2:1 MMR is the best candidate for the orbital resonance of these planets.

For our fits to \textit{b}, \textit{c} and the candidate \textit{d}, the period ratios between \textit{b} and \textit{c} are all consistent with a 2:1 MMR, excluding the 3:2 and 4:3 MMRs. However, while in the coplanar case the period ratio between \textit{b} and the candidate \textit{d} is consistent with a 2:1 MMR the stable and $N$-body cases are more consistent with a 3:1 MMR. The coplanar case is also consistent with a 3:2 MMR within $2\sigma$.

\begin{table*}
    \caption{Period ratios between the PDS 70 planets. The period ratios from the fits using only astrometry from PDS 70 \textit{b} and \textit{c} are shown in the first row. The remaining rows show the period ratios between each pair of planets from the fits where astrometry from \textit{b}, \textit{c} and the candidate \textit{d} is used. For each fit the median value of the posterior is shown, with subscripts and superscripts listing the 68\% and 95\% CIs (with the 95\% CIs in parentheses).}
    \label{tab:periods}
    \centering
    \begin{tabular}{c|c|c|c|c}
        \hline
        \hline
        Period Ratio & Astrometry & Coplanar & Stable & Stable (incl. $N$-body) \\
        \hline
        \textit{c}:\textit{b} & \textit{b}, \textit{c} & $1.99^{+0.15 (+0.29)}_{-0.13 (-0.26)}$ & $2.02^{+0.15 (+0.29)}_{-0.13 (-0.25)}$ & $2.08^{+0.13 (+0.22)}_{-0.12 (-0.23)}$ \\
        \hline
        \textit{c}:\textit{b} & \textit{b}, \textit{c}, \textit{d} & $2.11^{+0.13 (+0.28)}_{-0.13 (-0.24)}$ & $2.17^{+0.13 (+0.26)}_{-0.12 (-0.25)}$ & $2.16^{+0.09 (+0.21)}_{-0.05 (-0.23)}$ \\
        \textit{b}:\textit{d} & \textit{b}, \textit{c}, \textit{d} & $1.90^{+0.41 (+0.91)}_{-0.32 (-0.56)}$ & $2.7^{+0.6 (+1.2)}_{-0.5 (-0.8)}$ & $2.7^{+0.5 (+1.3)}_{-0.6 (-0.6)}$ \\
        \textit{c}:\textit{d} & \textit{b}, \textit{c}, \textit{d} & $4.0^{+0.9 (+1.9)}_{-0.7 (-1.3)}$ & $5.8^{+1.3 (+2.6)}_{-1.0 (-1.7)}$ & $6.1^{+1.2 (+2.3)}_{-1.5 (-1.6)}$ \\
        \hline
    \end{tabular}
\end{table*}

\section{Comparison to Evolutionary Models}
\label{sec:evo_models}

We used luminosity data from the \citet{Spiegel_2012} evolutionary models to differentiate between hot-, warm-, and cold-start scenarios for the PDS 70 planets. These models give the flux density at $d=10$~pc, from 0.8 to 15.0~\mum, for planets of mass $\Mp=1$--15~$M_{\rm Jup}$ for initial, i.e., post-formation, entropies $S_i=8$--13~\Sunits, for ages of $t=1$--1000~Myr. To obtain the bolometric luminosity of each model, we integrated the flux and multiplied by $4\pi d^2$ to convert from flux to luminosity. 

Entropy is defined physically only up to an additive constant. Naturally, we use the entropy zero-point adopted by \citet{Spiegel_2012}, who used the version of the \citet{Saumon_1995} equation of state that is offset by $\Delta S = 0.52$~\Sunits compared to the published tables (see \citealt{Marleau_2014} and Footnote~2 of \citealt{Mordasini_2017}). This applies also to the values in \citet{Burrows_1997}, \citet{Marley_2007}, \citet{Molliere_2012}, or \citet{Mordasini_2017}, while the \verb<MESA< code (e.g., \citealp{Paxton_2011,Jermyn_2023}), \citet{Berardo_2017}, or \citet{Marleau_2014} use the published \citet{Saumon_1995} tables. The upshot of the offset is that, at a given mass, a \citet{Spiegel_2012} model with (current) entropy $S$ will have the same luminosity as a \verb<MESA< model of entropy $S+\Delta S$.

\begin{figure*}
    \centering
    \includegraphics[width=0.9\linewidth]{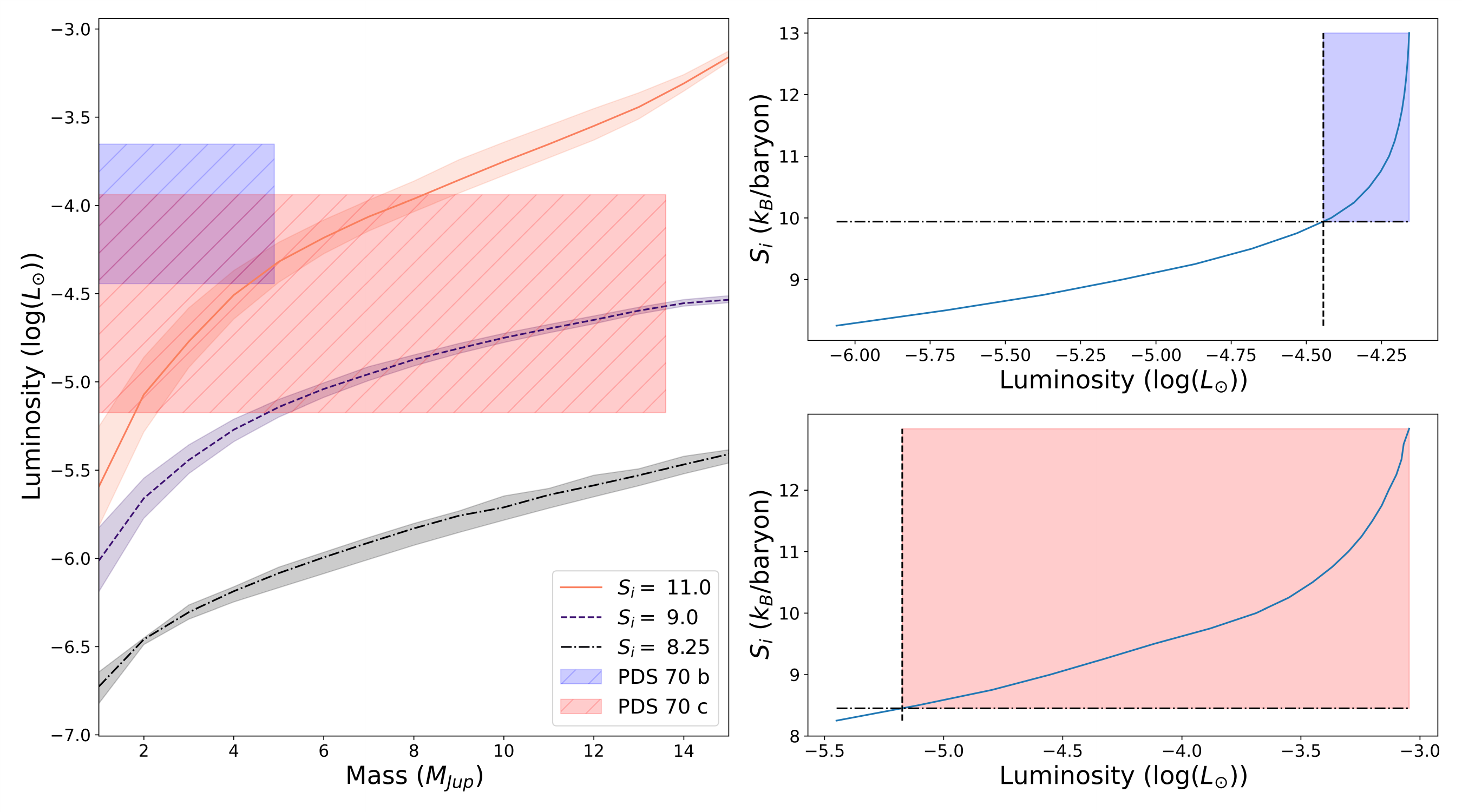}
    \caption{Comparison of the upper mass limits for PDS 70 \textit{b} and PDS 70 \textit{c} from the two-planet orbital parameter fit to cold-, warm-, and hot-start evolutionary models from \citet{Spiegel_2012}. The mass-luminosity curves for the different models, taken at the PDS 70 system age of 5.4 Myr are shown in the left panel, with their corresponding post-formation entropies given in the legend. The ribbon around each curve corresponds to the uncertainty in age of PDS 70. The blue and red boxes in the left panel show the possible masses and luminosities for \textit{b} and \textit{c} respectively, with luminosities taken from \citet{Wang_2020}. The upper right panel shows the luminosity-$S_i$ curve at the upper mass limit for \textit{b}. The blue shaded area in this panel, bounded by the dashed and dashed-dotted lines shows the possible range of $S_i$ for \textit{b} from this orbital parameter fit. The lower right panel is the same as upper right panel, but for \textit{c}.}
    \label{fig:evo_comp_2body}
\end{figure*}

\begin{figure*}
    \centering
    \includegraphics[width=0.9\linewidth]{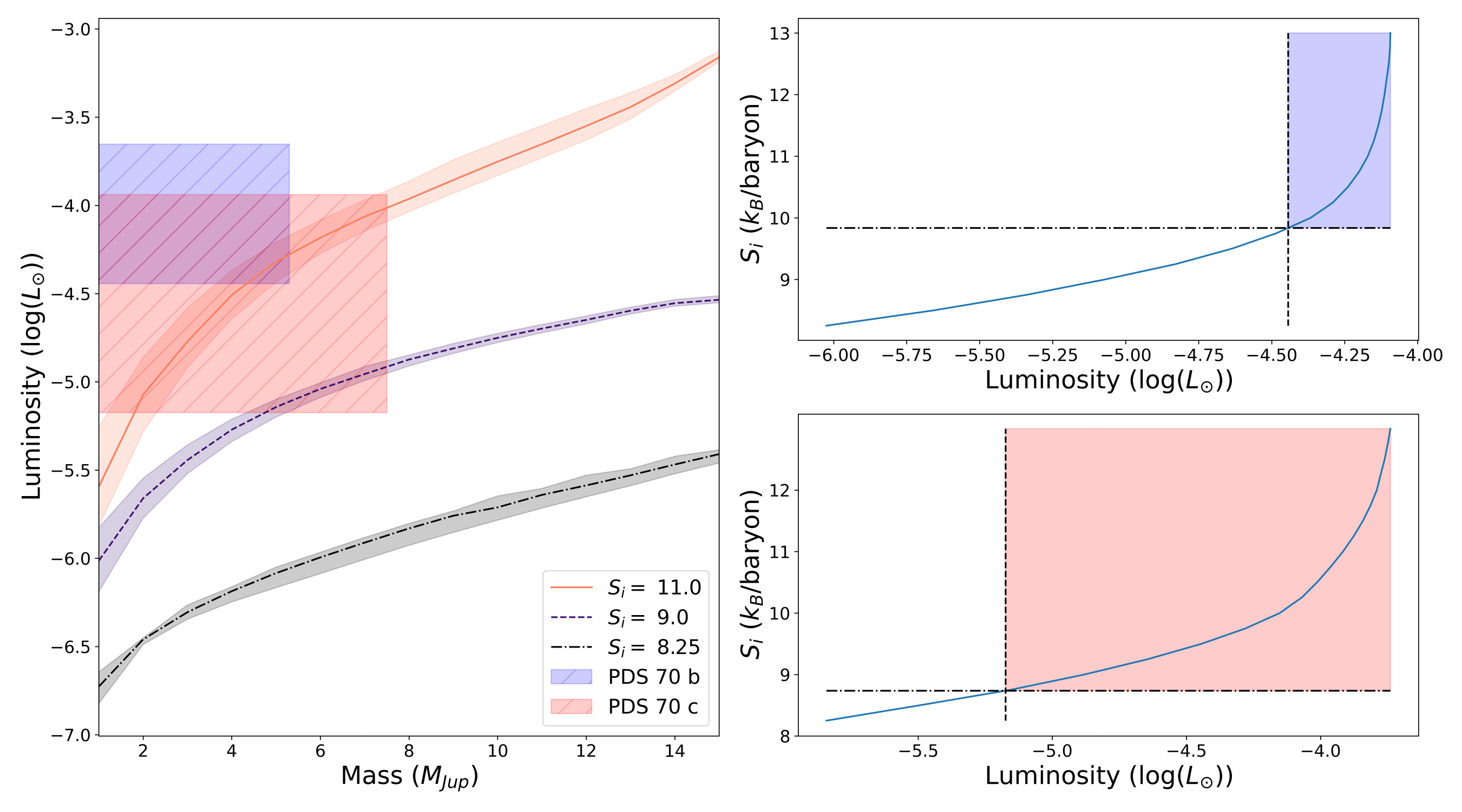}
    \caption{Same as Figure~\ref{fig:evo_comp_3body}, but with upper mass limits taken from the three-planet orbital parameter fit.}
    \label{fig:evo_comp_3body}
\end{figure*}

Figures~\ref{fig:evo_comp_2body} and~\ref{fig:evo_comp_3body} show example mass--luminosity curves at the PDS~70 system age of $5.4\pm1.0$ Myr (i.e., isochrones) for different values of $S_i$, from the \citet{Spiegel_2012} model. We calculated the luminosity at each mass value by linearly interpolating the age--luminosity relation for that mass at the system age. The shaded areas around the curves show the possible luminosity values given the uncertainty on the age of the system.

The blue boxes on the left panels of Figures~\ref{fig:evo_comp_2body} and~\ref{fig:evo_comp_3body} show the range of possible masses and luminosities for PDS 70 \textit{b} given the 95\% confidence upper mass limits from the 2-planet (Section~\ref{sec:2planets}) and 3-planet (Section~\ref{sec:3planets}) orbital parameter fits respectively. Similarly, the red boxes on the left panels of Figures~\ref{fig:evo_comp_2body} and~\ref{fig:evo_comp_3body} show the range of possible masses and luminosities for PDS 70 \textit{c} given the upper mass limit from each fit. Luminosities are taken from Tables~3 and~4 of \citet{Wang_2020}. Since these tables each contain four different sets of luminosity bounds (from four different spectral energy distribution models), the lowest and the highest bounds of the four are used as the low and high bounds for the planet luminosity in this paper.

From the left-hand panels of Figures~\ref{fig:evo_comp_2body} and~\ref{fig:evo_comp_3body} it is apparent that the coldest post-formation entropies are inconsistent with the observed luminosities of PDS 70 \textit{b} and \textit{c} given the mass constraints from both the two-planet and three-planet fits.

We used the upper mass limits for these planets, determined in Section~\ref{sec:orbital_dynamics}, to calculate the minimum values of $S_i$ necessary for their formation. The right-hand panels of Figures~\ref{fig:evo_comp_2body} and~\ref{fig:evo_comp_3body} show the luminosity-$S_i$ relation for each planet at their upper mass limit. Since luminosity monotonically increases with $S_i$, the minimum value of $S_i$ for the planet occurs where luminosity hits the lower bound for the planet. The luminosity-$S_i$ relation is given assuming the lower bound of $4.4$ Myr for the system age, since this is where the model is brightest within the age range for PDS 70.

For PDS 70 \textit{b} we calculated a lower limit for $S_i$ of 9.9 \Sunits using the upper mass limit from the two-planet fit and 9.8 \Sunits using the upper mass limit from the three-planet fit. For PDS 70 \textit{c} we calculated a lower limit of 8.5 \Sunits using the upper mass limit from the two-planet fit and 8.7 \Sunits using the upper mass limit from the three-planet fit.

\section{Discussion}
\label{sec:discussion}

\subsection{The Nature of PDS 70 \textit{d}}

From the currently available data the nature of PDS 70 \textit{d} is unclear. It is located close to the outer edge of the inner disk and has been consistently detected with multiple telescopes and instruments over multiple epochs \citep{Mesa_2019, Christiaens_2024, Hammond_2025}. However, its spectrum is more similar to star light reflected by dust than thermal emission by a planet \citep{Mesa_2019}. This suggests two possibilities:
\begin{enumerate}
    \item PDS 70 \textit{d} is not a planet and is instead a disk feature, such as a dust clump
    \item PDS 70 \textit{d} is a planet embedded in a dust envelope
\end{enumerate}

When \textit{d} is included in the analysis the fraction of orbital configurations deemed to be stable after $N$-body analysis is small. The $N$-body analysis also indicates that \textit{d} must have a comparatively low mass in order for the PDS 70 planetary system to remain stable. Either $N$-body analysis is very efficient at constraining the orbital parameter space for this system or there is a lack of stable orbit configuration for a three-body system. The latter case would suggest that \textit{d} is a disk feature and not a planet.

It is important to note that we have not included the impacts of gas drag in our $N$-body analyses. While this is not important for PDS 70 \textit{b} and \textit{c}, since they sit comfortably inside the gap in the disk, the candidate \textit{d} sits close to the outer edge of the inner disk. On short timescales planet-disk interaction dampens the eccentricity and inclination of the planet \citep{Goldreich_2003, Tanaka_2004}, which could therefore enhance the stability of the system.

\subsection{Viability of Cold-Start Formation}

Our analysis in Section~\ref{sec:evo_models} adds PDS 70 \textit{b} and \textit{c} to the list of directly-imaged exoplanets that are inconsistent with cold-start formation. \citet{Marleau_2014} found $S_i > 9.2$ \Sunits for 2M1207~b and the planets in the HR~8799 system, as well as $S_i > 10.5$ for $\beta$~Pic~b. \citet{Brandt_2021a} used dynamical mass measurements of $\beta$~Pic~c to determine that it is inconsistent with the cold-start models of \citet{Spiegel_2012}. While this may suggest that true cold-start giant planets are rare (or even non-existent), we cannot rule out observational bias. Given a constant mass and age, higher $S_i$ leads to higher luminosity, and therefore these planets are more likely to be detected via direct-imaging surveys.

\subsection{Are the PDS 70 Planets Still Accreting?}

In order to compare PDS 70 \textit{b} and \textit{c} to the \cite{Spiegel_2012} evolutionary models, we have made the assumption that these planets are "done" accreting gas. \cite{Haffert_2019} gives relatively low accretion rates of $\sim 0.01$ $M_{\rm Jup}$ Myr$^{-1}$ for both planets, based on their H$\alpha$ emission. However, modelling of the spectral energy distribution of PDS 70 c by \cite{Choksi_2025} suggests that its accretion rate could be as high as $0.6$ $M_{\rm Jup}$ Myr$^{-1}$. This would result in the accretion of multiple $M_{\rm Jup}$ of gas before the dispersal of the protoplanetary disk, significantly increasing the luminosity of the planet and affecting the minimum $S_i$ calculation.

Even if we assume that the accretion rates for these planets are low, our analysis in Section~\ref{sec:evo_models} also assumes that the planets are the same age as the star. If the planets formed more recently than the star, then their luminosity and mass will be consistent with lower $S_i$ evolutionary tracks. To understand how this impacts our minimum $S_i$ values, we performed the same analysis as in Section~\ref{sec:evo_models}, but assuming a lower limit of $1$ Myr on the age. For this analysis we used the mass limits given from the orbit fit to \textit{b} and \textit{c} only. In this case, we find a minimum $S_i$ of 9.5 \Sunits for \textit{b} and 8.4 \Sunits for \textit{c}. Therefore, even if these planets are newly formed, they are still inconsistent with the coldest-start models.

Planet-planet collisions could also provide an additional source of heat, which is not accounted for in our determination of the minimum values of $S_i$. However, for giant planets of multiple Jupiter mass, planet--planet scattering tends to lead to planet ejection rather than collision \citep{Rasio_1996, Weidenschilling_1996}.

\section{Conclusions}
\label{sec:conclusions}

Using new relative astrometry for PDS 70 \textit{b} and \textit{c} obtained with VLTI/GRAVITY we obtained updated orbital parameters for these planets, including dynamical mass estimates. Fitting to astrometry from \textit{b} and \textit{c} only, we calculated an upper mass limits (with 95\% confidence) of 4.9 $M_{\rm Jup}$ for \textit{b} and 13.6 $M_{\rm Jup}$ for \textit{c}.

It is unclear whether PDS 70 \textit{d} is a planet or a disk feature. Our $N$-body stability analysis of orbits fit to astrometry of \textit{b}, \textit{c} and the candidate \textit{d} finds a relative lack of stable orbital configurations. However, we do not consider gas drag, which may lead to more stable orbits. From our orbital parameter fits that include \textit{d}, we find 95\% confidence upper mass limits of 5.3 $M_{\rm Jup}$, 7.5 $M_{\rm Jup}$, and 2.2 $M_{\rm Jup}$ for \textit{b}, \textit{c} and the candidate \textit{d} respectively.

In all of our analyses the period ratio of PDS 70 \textit{c} to \textit{b} is consistent with a 2:1 MMR, while ruling out the 3:2 and 4:3 MMRs. When we include the candidate \textit{d} and enforce coplanarity the period ratio of \textit{b} to \textit{d} is consistent with the 2:1 and 3:2 MMRs. However, when we enforce stability and when we employ $N$-body stability analysis the period ratio becomes inconsistent with both of these MMRs, and instead becomes more consistent with the 3:1 MMR.

We eliminate the possibility of a cold-start for both PDS 70 \textit{b} and PDS 70 \textit{c} through comparison of our dynamical mass estimates to hot-, warm- and cold-start evolutionary models. This is true of both the two-planet and three-planet fits. We determine a minimum $S_i$ of 9.9 \Sunits for \textit{b} using the two-planet fit, and of 9.8 \Sunits using the three-planet fit. Similarly, we determine a minimum $S_i$ of 8.5 \Sunits for \textit{c} using the two-planet fit and 8.7 \Sunits using the three-planet fit. Even in the case where these planets are newly-formed they are both inconsistent with the coldest-start models, with minimum $S_i$ values of 9.5 \Sunits and 8.4 \Sunits respectively.

\begin{acknowledgements}
Based on observations collected at the European Organisation for Astronomical Research in the Southern Hemisphere under ESO programmes 1104.C-0651(A), 1104.C-0651(C) and 105.209D.001. We thank Zhouijan Zhang for useful discussions. 
I.H. acknowledges a Research Training Program scholarship from the Australian government.
R.B. acknowledges the financial support from DFG under Germany’s Excellence Strategy EXC 2181/1-390900948, Exploratory project EP 8.4 (the Heidelberg STRUCTURES Excellence
	Cluster). J.S.-B. acknowledges the support received from the UNAM PAPIIT project IA 105023.
G.-D.M.\ acknowledges the support from the European Research Council (ERC) under the Horizon 2020 Framework Program via the ERC Advanced Grant ``ORIGINS'' (PI: Henning), Nr.~832428,
and via the research and innovation programme ``PROTOPLANETS'', grant agreement Nr.~101002188 (PI: Benisty).
\end{acknowledgements}

\bibliographystyle{yahapj}
\bibliography{ref}


\begin{appendix}

\renewcommand{\arraystretch}{1.1}

\section{New reduction of VLT/SINFONI data}
\label{sec:SINFONI}

\citet{Christiaens2019} reported the astrometry of planet \textit{b} detected in VLT/SINFONI data acquired on 10 May 2014. For planet \textit{c} the earliest reported astrometry is from a SPHERE dataset of 2018 \citep{Mesa_2019}. Nonetheless the new reduction of the SINFONI dataset presented in \citet{Hammond_2025} not only revealed candidate \textit{d}, but also provided a clear detection of planet \textit{c}. This paragraph summarizes how we calculated the astrometry for planet \textit{c} from this dataset.

Details on the data calibration and processing can be found in \citet{Christiaens2019} and \citet{Hammond_2025}. For stellar PSF modeling and subtraction we considered principal component analysis \citep[PCA; ][]{Soummer2012, Amara2012} to leverage in two consecutive steps both the spectral \citep[SDI;][]{SparksFord2002} and the angular differential imaging strategies \citep[ADI;][]{Marois2006}. We used the 2-step PCA implementation in VIP \citep{GomezGonzalez2017, Christiaens2023}. We collapsed only the spectral channels longward of 2.3 $\mu$m during PCA-SDI, as these benefit from the highest Strehl ratio and the highest flux for planet \textit{c}. For PCA-ADI, we considered a 1-FWHM rotation threshold at the separation of planet c when building the PCA library for each residual frame resulting from PCA-SDI.

Since the signals from planet \textit{c} are merged with the signals from the illuminated edge of the outer disk, estimating its astrometry using classical point-source forward-modelling techniques would lead to biased estimates. Instead we first isolated the signal of \textit{c} from the disk by subtracting a disk model. The disk model was built from the south side of the disk, leveraging the expected symmetry with respect to the semi-minor axis. In practice, we (i) rotated the image by an angle corresponding to $270\degr-$PA$_{\beta}$, where PA$_{\beta}$ is the semi-minor axis of the disk PA$_{\beta}$ pointing to the near side of the disk, (ii) subtracted from each north or south half of the image the signals from the other half, and (iii) rotated back the image by PA$_{\beta}-270\degr$. A similar strategy as the direct negative fake companion technique presented in \citet{Christiaens_2024} was then adopted to retrieve the astrometry of \textit{c}.

\begin{figure}
    \centering
    \includegraphics[width=\linewidth]{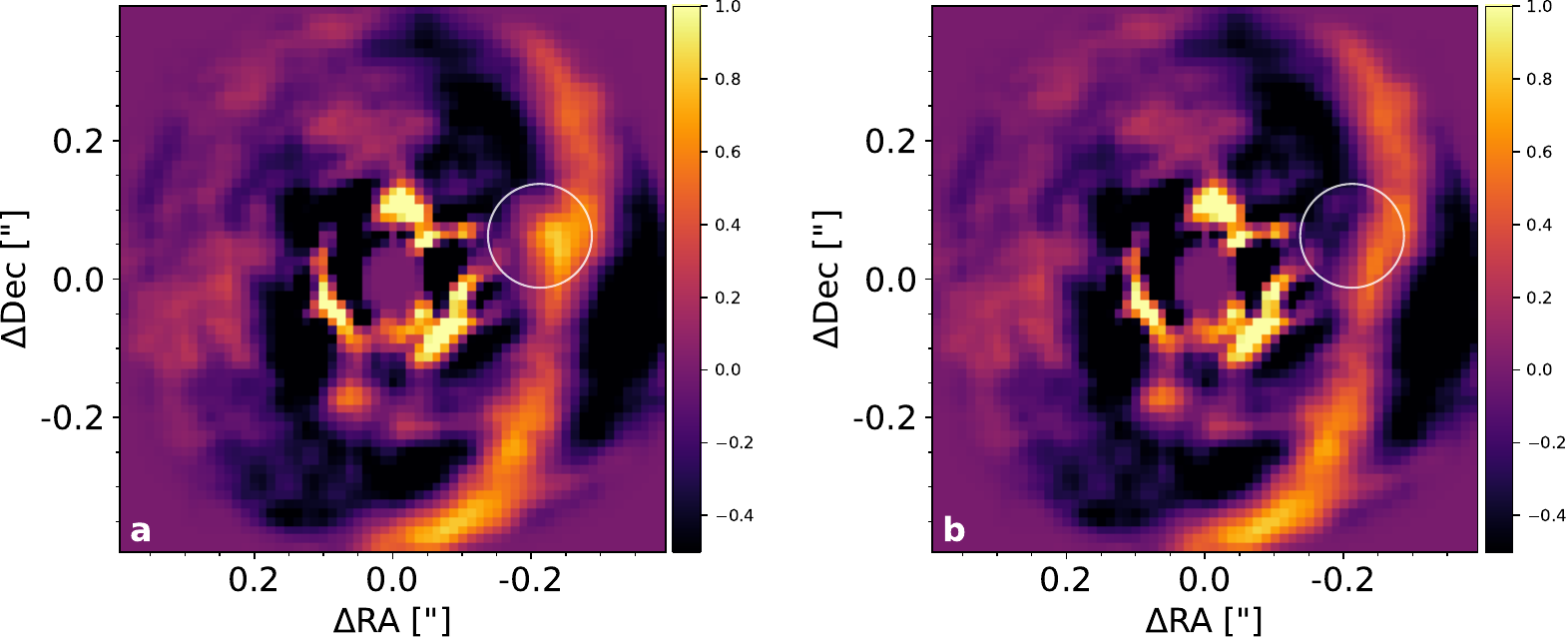}
    \caption{Final image obtained from our new reduction of the 2014-05-10 SINFONI data, before (a) and after (b) subtracting the estimated contribution of planet \textit{c} as explained in Sec.~\ref{sec:SINFONI}.}
    \label{fig:c_in_SINFONI}
\end{figure}

The dominant source of uncertainty in this procedure is associated with the presence of faint residual extended signals similar to the spiral accretion stream reported in \citet{Christiaens_2024}, henceforth the figure of merit to be considered for optimal estimation of the astrometry of \textit{c}. We therefore considered both a range of values for PA$_{\beta}$ (252\degr~to 260\degr~per step of 0.5\degr), and two different figures of merit: (i) minimization of the absolute value of the determinant of the Hessian matrix at the estimated location of \textit{c}, and (ii) minimization of the sum of absolute values of residual pixel intensities within an aperture at the esimated location of \textit{c}. Two sub-cases were considered for (i) and (ii): a Hessian matrix of either 1$\times$1 or 2$\times$2, and aperture sizes of 0.5 and 1 FWHM, respectively. This made for 68 estimates, which were visually vetted and considered plausible based on the shape of the outer disk edge after subtraction of \textit{c}. The median value of these 68 astrometric estimates is reported in Table~\ref{tab:bcd_lit_astrometry}.
The subtraction of \textit{c} with the optimal parameters found with this approach is shown in Figure~\ref{fig:c_in_SINFONI}.

We considered the standard deviation of these 68 estimates as the uncertainty associated with the disk. As final uncertainties, we conservatively combined in quadrature (i) the uncertainty associated with the disk, (ii) the residual speckle noise uncertainties - assumed to be the same as reported for planet b in \citet{Christiaens2019}, and (iii) systematic uncertainties on the plate scale and true north conservatively assumed to be 0.4~mas and 0.5~\degr, respectively \citep{Christiaens2019}. These final uncertainties are also reported in Table~\ref{tab:bcd_lit_astrometry}. We did not attempt to re-estimate the residual speckle noise uncertainty associated to \textit{c}, due to the presence of bright and extended signals at that radius. Since the estimate for \textit{b} was obtained after similar data processing, the latter can be considered as a conservative estimate of the residual speckle uncertainty, as \textit{b} is slightly more inward than \textit{c}.

\section{Archival Data}
\label{sec:lit_data}
Table~\ref{tab:bcd_lit_astrometry} collects the existing astrometry for both confirmed planets and for candidate \textit{d} of the PDS 70 system.
    
\onecolumn
\begin{table*}[h]
  \caption{Literature astrometry for PDS 70 \textit{b}, \textit{c} and \textit{d}.} 
  \label{tab:bcd_lit_astrometry} 
  \centering
  \begin{tabular}{c|c c c|c c c c|c c c c|c} 
    \hline 
    & Date & MJD & Instrument & $\Delta$R.A. & $\sigma_{\Delta {\rm R.A.}}$ & $\Delta$Dec. & $\sigma_{\Delta {\rm Decl.}}$ & Sep. & $\sigma_{\rm Sep.}$ & P.A. & $\sigma_{\rm P.A.}$ & Ref. \\
    & & (days) & & (mas) & (mas) & (mas) & (mas) & (mas) & (mas) & (deg) & (deg) & \\
    \hline 
    \textit{b} & 2012-03-31 & 56017 & NICI & 58.7 & 10.7 & $-182.7$ & 22.2 & 191.9 & 21.4 & 162.2 & 3.7 & [1] \\ 
    \textit{b} & 2014-05-10 & 56787 & SINFONI & 70.3 & 9.6 & $-180.3$ & 5.9 & 193.5 & 4.9 & 158.7 & 3.0 & [3] \\
    \textit{b} & 2015-05-03 & 57145 & IRDIS & 83.1 & 3.9 & $-173.5$ & 4.3 & 192.3 & 4.2 & 154.5 & 1.2 & [1] \\ 
    \textit{b} & 2015-05-03 & 57145 & IRDIS & 83.9 & 3.6 & $-178.5$ & 4.0 & 197.2 & 3.9 & 154.9 & 1.1 & [1] \\ 
    \textit{b} & 2015-06-01 & 57174 & IRDIS & 89.4 & 6.0 & $-178.3$ & 7.1 & 199.5 & 6.9 & 153.4 & 1.8 & [1] \\ 
    \textit{b} & 2015-06-01 & 57174 & IRDIS & 86.9 & 6.2 & $-174.0$ & 6.4 & 194.5 & 6.3 & 153.5 & 1.8 & [1] \\ 
    \textit{b} & 2016-05-14 & 57522 & IRDIS & 90.2 & 7.3 & $-170.8$ & 8.6 & 193.2 & 8.3 & 152.2 & 2.3 & [1] \\ 
    \textit{b} & 2016-05-14 & 57522 & IRDIS & 95.2 & 4.8 & $-175.0$ & 7.7 & 199.2 & 7.1 & 151.5 & 1.6 & [1] \\ 
    \textit{b} & 2016-06-01 & 57540 & NaCo & 94.5 & 22.0 & $-164.4$ & 27.6 & 189.6 & 26.3 & 150.6 & 7.1 & [1] \\ 
    \textit{b} & 2018-02-24 & 58174 & IRDIS & 109.6 & 7.9 & $-157.7$ & 7.9 & 192.1 & 7.9 & 147.0 & 2.4 & [2] \\ 
    \textit{b} & 2018-02-24 & 58174 & IRDIS & 110.0 & 7.9 & $-157.6$ & 8.0 & 192.2 & 8.0 & 146.8 & 2.4 & [2] \\ 
    \textit{b} & 2018-06-20 & 58289 & MUSE & 96.8 & 25.4 & $-147.9$ & 25.9 & 176.8 & 25 & 146.8 & 8.5 & [4] \\ 
    \textit{b} & 2019-06-08 & 58642 & NIRC2 & 110.9 & 6.8 & $-136.4$ & 6.8 & 175.8 & 6.9 & 140.9 & 2.2 & [6]  \\ 
    \textit{b} & 2019-07-16 & 58680 & GRAVITY & 102.626 & 0.087 & $-139.879$ & 0.291 & 173.49 & 0.24 & 143.733 & 0.061 & [7] \\ 
    \textit{b} & 2020-02-10 & 58889 & GRAVITY & 104.545 & 0.229 & $-135.081$ & 0.077 & 170.81 & 0.15 & 142.262 & 0.063 & [7] \\ 
    \textit{b} & 2021-07-15 & 59410 & IRDIS & 118 & 11 & $-118$ & 11 & 167.5 & 11.0 & 135.0 & 4.0 & [10] \\
    \textit{b} & 2021-08-21 & 59447 & IRDIS & 108.5 & 8.2 & $-118.0$ & 8.2 & 160.3 & 8.0 & 137.4 & 3.0 & [10] \\
    \textit{b} & 2021-08-22 & 59448 & IRDIS & 110.8 & 5.9 & $-125.2$ & 5.9 & 167.2 & 6.0 & 138.5 & 2.0 & [10] \\
    \textit{b} & 2021-09-02 & 59459 & IRDIS & 109.8 & 5.9 & $-128.5$ & 6.0 & 169.0 & 6.0 & 139.5 & 2.0 & [10] \\
    \textit{b} & 2021-09-04 & 59461 & IRDIS & 110.1 & 5.9 & $-128.0$ & 6.0 & 168.9 & 6.0 & 139.3 & 2.0 & [10] \\
    \textit{b} & 2022-02-28 & 59638 & IRDIS & 114.3 & 5.2 & $-104.0$ & 5.2 & 154.6 & 5.0 & 132.3 & 2.0 & [10] \\
    \textit{b} & 2022-04-24 & 59693 & MagAO-X & 110.8 & 2.3 & $-112.8$ & 2.3 & 158.1 & 3.0 & 135.5 & 0.5 & [11] \\
    \textit{b} & 2023-02-24 & 59999 & NIRISS & 110.7 & 6.7 & $-97.9$ & 6.3 & 147.8 & 8.2 & 131.5 & 1.6 & [8] \\
    \textit{b} & 2023-03-08 & 60011 & MagAO-X & 116.7 & 2.4 & $-105.8$ & 2.3 & 157.5 & 3.0 & 132.18 & 0.5 & [11] \\
    \textit{b} & 2023-03-08 & 60011 & NIRCam & 128 & 11 &  $-104$  & 11 & 164.5 & 10.4 & 129.0 & 4.1 & [9] \\
    \textit{b} & 2024-03-25 & 60394 & MagAO-X & 115.0 & 2.4 & $-97.1$ & 2.2 & 150.5 & 3.0 & 130.18 & 0.50 & [11] \\
    \hline
    \textit{c} & 2014-05-10 & 56787 & SINFONI & $-212.1$ & 10.5 & 66.2 & 7.3 & 222.2 & 10.3 & 287.3 & 2.0 & [13] \\
    \textit{c} & 2018-02-25 & 58174 & IRDIS & $-205.0$ & 13.0 & 41.0 & 6.0 & 209 & 13 & 281.2 & 0.5 & [5] \\ 
    \textit{c} & 2018-06-20 & 58289 & MUSE & $-233.7$ & 26.7 & 28.7 & 25.0 & 235.5 & 25.0 & 277.0 & 6.5 & [4]  \\ 
    \textit{c} & 2019-03-06 & 58548 & IRDIS & $-222.0$ & 8.0 & 39.0 & 4.0 & 225 & 8 & 279.9 & 0.5 & [5] \\ 
    \textit{c} & 2019-06-08 & 58642 & NIRC2 & $-219.7$ & 7.8 & 40.3 & 8.0 & 223.4 & 8.0 & 280.4 & 2.0 & [6]  \\ 
    \textit{c} & 2019-07-19 & 58683 & GRAVITY & $-214.98$ & 0.296 & 32.323 & 0.396 & 217.40 & 0.30 & 278.55 & 0.10 & [7] \\ 
    \textit{c} & 2019-07-19 & 58683 & GRAVITY & $-214.929$ & 0.335 & 32.222 & 0.437 & 217.33 & 0.34 & 278.53 & 0.11 & [7] \\ 
    \textit{c} & 2020-02-10 & 58889 & GRAVITY & $-214.333$ & 0.144 & 27.212 & 0.249 & 216.05 & 0.15 & 277.236 & 0.066 & [7] \\ 
    \textit{c} & 2021-07-15 & 59410 & IRDIS & $-208.2$ & 5.0 & 12.4 & 3.6 & 208.6 & 5.0 & 273.4 & 1.0 & [10] \\
    \textit{c} & 2021-08-22 & 59448 & IRDIS & $-203.6$ & 8.0 & 7.5 & 7.1 & 203.7 & 8.0 & 272.1 & 2.0 & [10] \\
    \textit{c} & 2021-09-02 & 59459 & IRDIS & $-207.0$ & 6.0 & 11.0 & 7.2 & 207.3 & 6.0 & 273.2 & 2.0 & [10] \\
    \textit{c} & 2022-02-28 & 59638 & IRDIS & $-195.6$ & 9.0 & 19 & 10 & 196.5 & 9.0 & 275.5 & 3.0 & [10] \\
    \textit{c} & 2023-02-24 & 59999 & NIRISS & $-221.1$ & 8.7 & $-1.5$ & 5.4 & 221.1 & 8.7 & 269.6 & 1.4 & [8] \\
    \textit{c} & 2023-03-08 & 60011 & MagAO-X & $-206.5$ & 1.0 & 0.0 & 1.0 & 206.5 & 1.0 & 270.0 & 0.25 & [11] \\
    \textit{c} & 2023-03-08 & 60011 & NIRCam & $-196$ & 23 &  $-1$  & 23 & 195.7 & 23.3 & 269.7 & 6.8 & [9] \\
    \textit{c} & 2024-03-25 & 60394 & MagAO-X & $-206.4$ & 3.0 & $-7.2$ & 1.8 & 206.55 & 3.0 & 268.0 & 0.50 & [11] \\
    \hline
    \textit{d} & 2014-05-10 & 56787 & SINFONI & $-23.0$ & 9.2 & 114.9 & 12.4 & 117.2 & 12.5 & 348.7 & 4.4 & [12] \\ 
    \textit{d} & 2016-06-01 & 57540 & IFS & $-49.0$ & 22.3 & 103.3 & 18.4 & 114.3 & 17.1 & 334.6 & 11.7 & [12] \\ 
    \textit{d} & 2018-02-25 & 58174 & IFS & $-79.3$ & 5.8 & 87.2 & 5.8 & 117.9 & 6.0 & 317.7 & 2.7 & [12] \\
    \textit{d} & 2019-03-06 & 58548 & IFS & $-84.1$ & 7.3 & 86.2 & 7.3 & 120.4 & 8.3 & 315.7 & 2.9 & [12] \\
    \textit{d} & 2021-05-17 & 59351 & IFS & $-86.3$ & 11.5 & 55.2 & 10.7 & 102.4 & 12.1 & 302.6 & 5.6 & [12] \\ 
    \textit{d} & 2021-08-21 & 59447 & IFS & $-89.9$ & 12.7 & 64.6 & 12.6 & 110.7 & 12.7 & 305.7 & 6.5 & [12] \\ 
    \textit{d} & 2021-09-03 & 59460 & IFS & $-92.4$ & 10.9 & 52.7 & 11.3 & 106.4 & 10.7 & 299.7 & 6.2 & [12] \\ 
    \textit{d} & 2022-02-28 & 59638 & IFS & $-95.2$ & 6.7 & 53.7 & 7.2 & 109.3 & 6.5 & 299.4 & 3.9 & [12] \\ 
    \textit{d} & 2023-03-08 & 60011 & NIRCam & $-95$ & 23 & 40 & 23 & 103.4 & 23.2 & 293.0 & 12.7 & [9] \\ 
    \hline
  \end{tabular}
  \tablefoot{References: [1] \cite{Keppler_2018}; [2] \cite{Mueller_2018}; [3] \cite{Christiaens2019}; [4] \cite{Haffert_2019}; [5] \cite{Mesa_2019}; [6] \cite{Wang_2020}; [7] \cite{Wang_2021}; [8] \cite{Blakely_2024}; [9] \cite{Christiaens_2024}; [10] \cite{Wahhaj2024}; [11] \cite{Close2025}; [12] \cite{Hammond_2025}; [13] This work.}
\end{table*}
\twocolumn


\section{Corner Plots of Planet Orbital Parameters}
\label{sec:corner_plots}

\begin{figure*}
    \centering
    \includegraphics[width=0.9\linewidth]{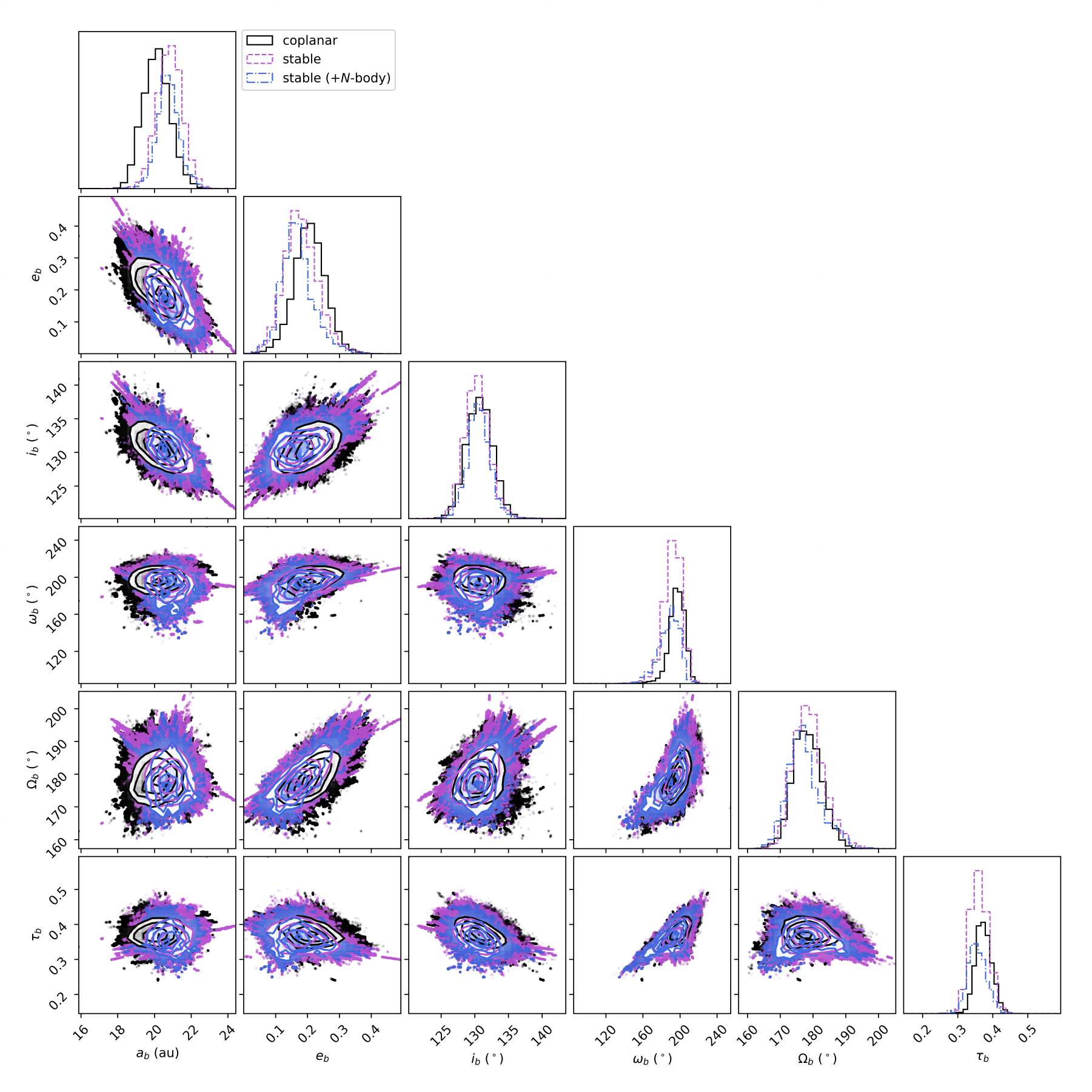}
    \caption{Corner plot of the posteriors for the orbital parameters of PDS 70 \textit{b} from all MCMC orbital parameter fits to astrometry of PDS 70 \textit{b} and \textit{c} only. Posteriors from the fit assuming coplanarity are shown in black (solid lines). Posteriors from the fit assuming stability are shown both before (purple, dashed lines) and after (blue, dashed and dotted lines) $N$-body stability analysis.}
    \label{fig:corner_bc_b}
\end{figure*}

\begin{figure*}
    \centering
    \includegraphics[width=0.9\linewidth]{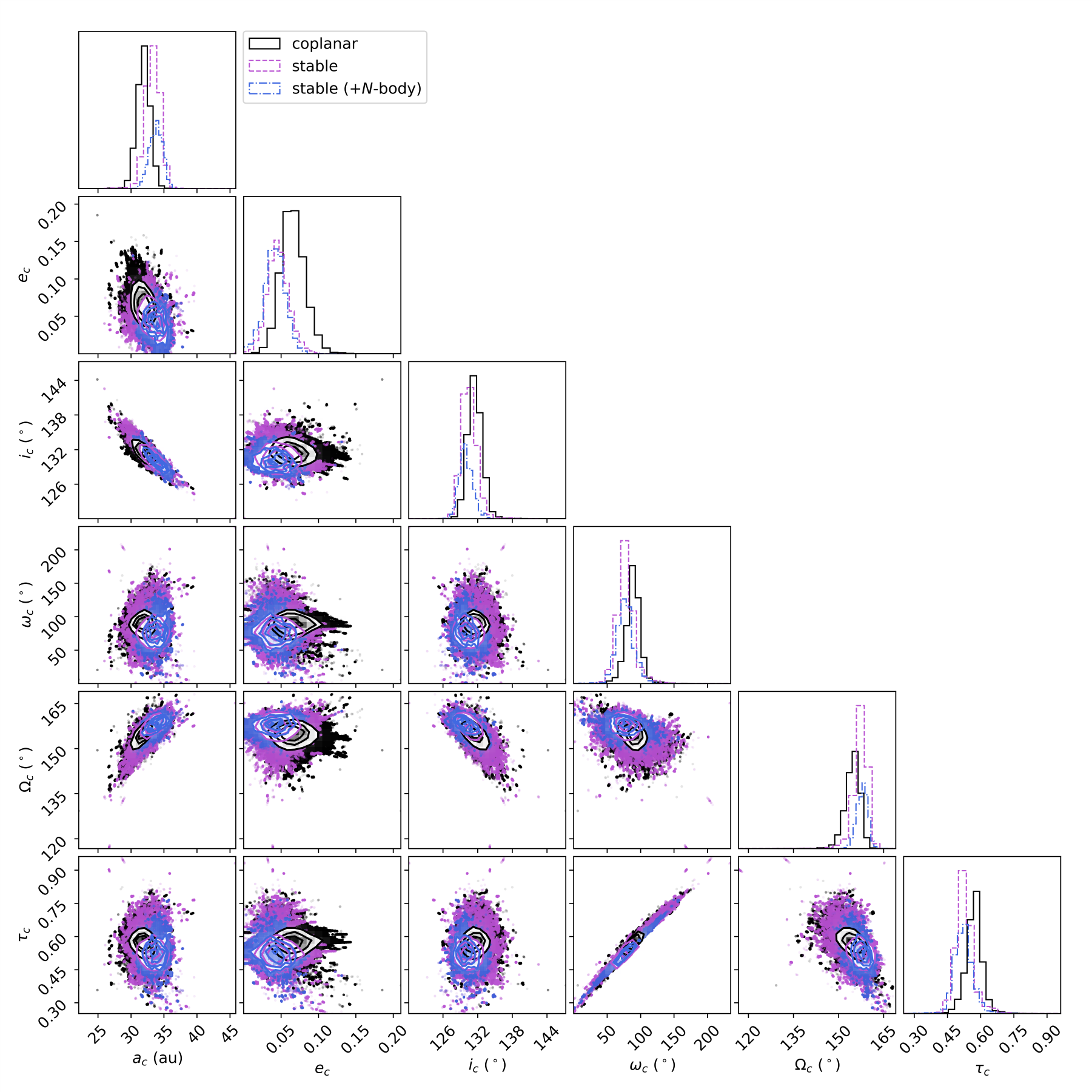}
    \caption{Corner plot of the posteriors for the orbital parameters of PDS 70 \textit{c} from all MCMC orbital parameter fits to astrometry of PDS 70 \textit{b} and \textit{c} only. Posteriors from the fit assuming coplanarity are shown in black (solid lines). Posteriors from the fit assuming stability are shown both before (purple, dashed lines) and after (blue, dashed and dotted lines) $N$-body stability analysis.}
    \label{fig:corner_bc_c}
\end{figure*}

\begin{figure*}
    \centering
    \includegraphics[width=0.9\linewidth]{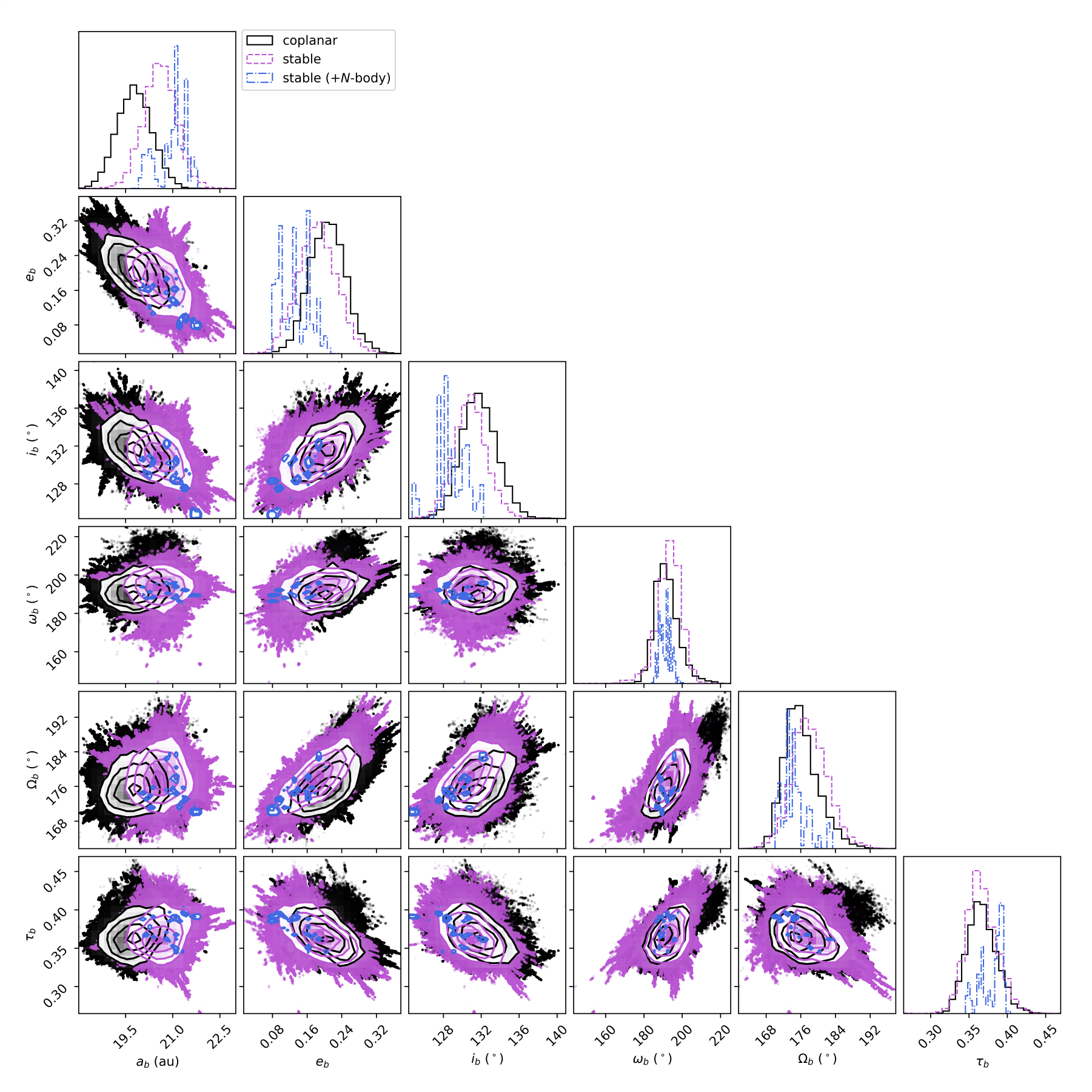}
    \caption{Corner plot of the posteriors for the orbital parameters of PDS 70 \textit{b} from all MCMC orbital parameter fits to astrometry of PDS 70 \textit{b}, \textit{c} and \textit{d}. Posteriors from the fit assuming coplanarity are shown in black (solid lines). Posteriors from the fit assuming stability are shown both before (purple, dashed lines) and after (blue, dashed and dotted lines) $N$-body stability analysis.}
    \label{fig:corner_bcd_b}
\end{figure*}

\begin{figure*}
    \centering
    \includegraphics[width=0.9\linewidth]{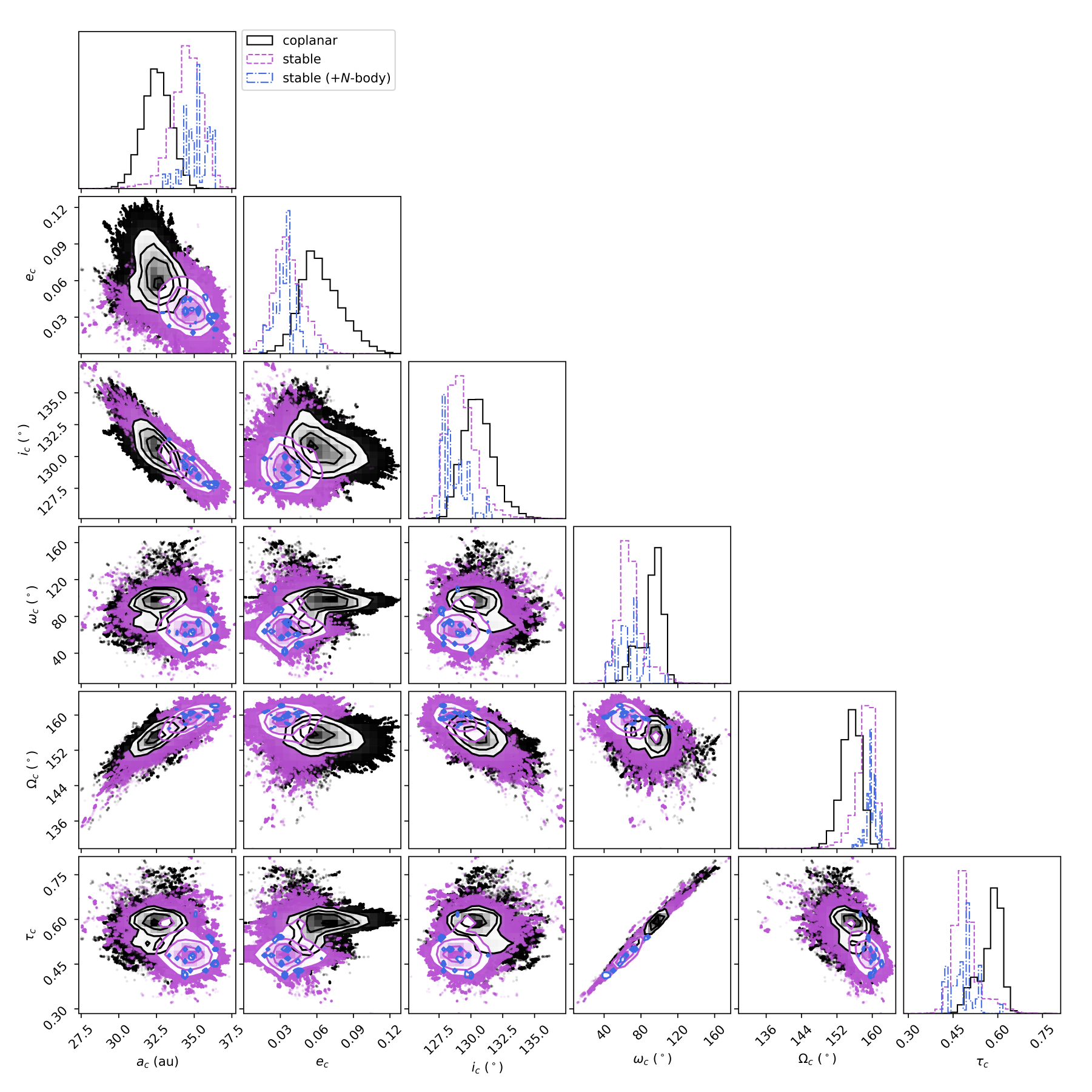}
    \caption{Corner plot of the posteriors for the orbital parameters of PDS 70 \textit{c} from all MCMC orbital parameter fits to astrometry of PDS 70 \textit{b}, \textit{c} and \textit{d}. Posteriors from the fit assuming coplanarity are shown in black (solid lines). Posteriors from the fit assuming stability are shown both before (purple, dashed lines) and after (blue, dashed and dotted lines) $N$-body stability analysis.}
    \label{fig:corner_bcd_c}
\end{figure*}

\begin{figure*}
    \centering
    \includegraphics[width=0.9\linewidth]{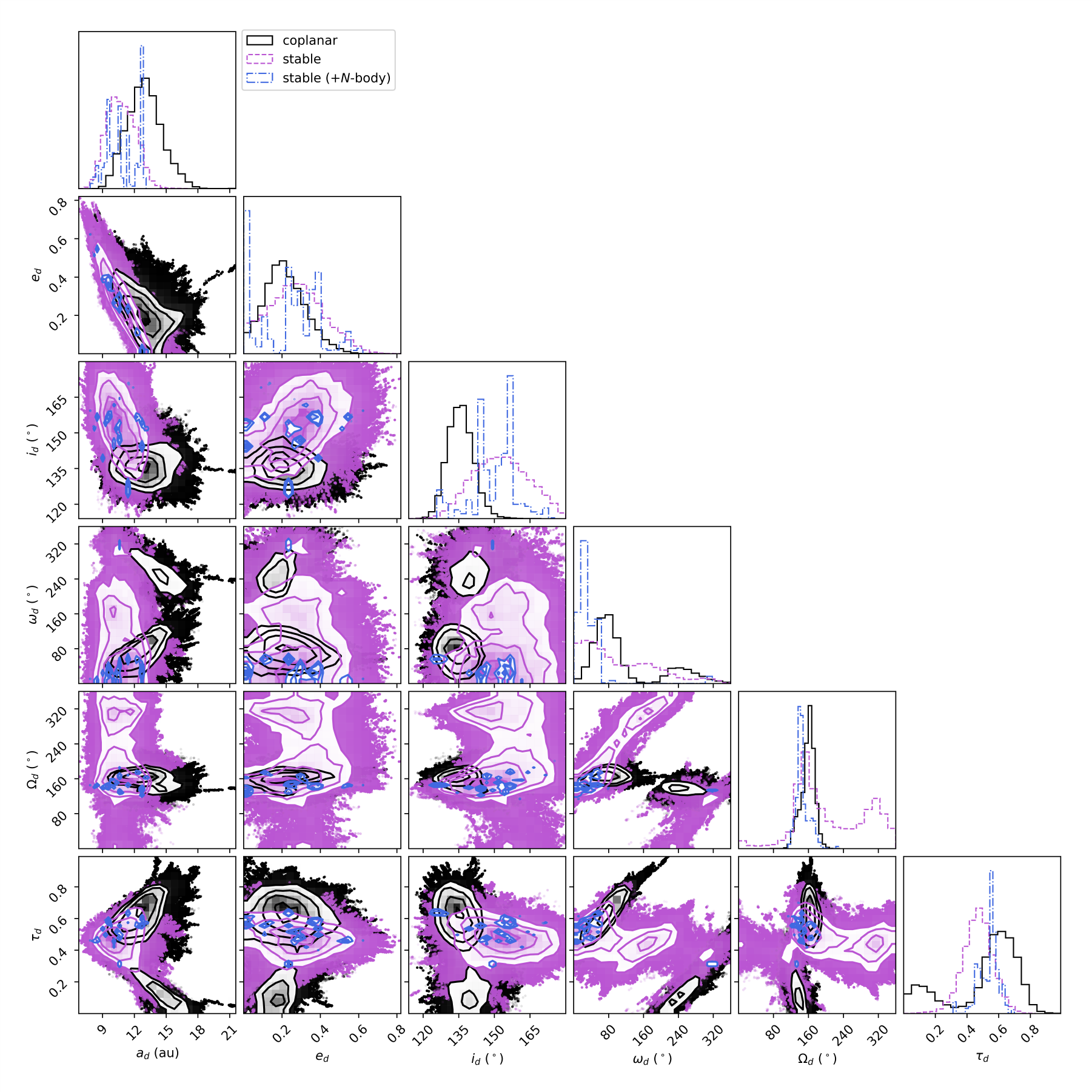}
    \caption{Corner plot of the posteriors for the orbital parameters of PDS 70 \textit{d} from all MCMC orbital parameter fits to astrometry of PDS 70 \textit{b}, \textit{c} and \textit{d}. Posteriors from the fit assuming coplanarity are shown in black (solid lines). Posteriors from the fit assuming stability are shown both before (purple, dashed lines) and after (blue, dashed and dotted lines) $N$-body stability analysis.}
    \label{fig:corner_bcd_d}
\end{figure*}

\end{appendix}

\end{document}